\documentclass[a4paper,fleqn]{cas-sc}
\usepackage[section]{placeins}
\usepackage{xcolor}
\usepackage{graphicx} 
\usepackage{amsmath}
\usepackage{amssymb}
\usepackage{subfig}
\usepackage{siunitx}
\usepackage{cleveref}
\usepackage{nicefrac}

\usepackage[numbers]{natbib}
\newcommand {\spec}[1]{^{(#1)}}

\newcommand{\fs}{f\spec{\alpha}}
\newcommand{\fst}{f\spec{\alpha,t}}
\newcommand{\Ts}{T\spec{\alpha}}
\newcommand{\Tsrel}{T\spec{\alpha,\mathrm{rel}}}

\newcommand{\us}{\mathbf u\spec{\alpha}}
\newcommand{\usrel}{\mathbf u\spec{\alpha,\mathrm{rel}}}
\newcommand{\uhatsrel}{\hat{\mathbf u}\spec{\alpha,\mathrm{rel}}}
\newcommand{\rhos}{\rho\spec{\alpha}}
\newcommand{\ns}{n\spec{\alpha}}

\newcommand{\sigmas}{\sigma\spec{\alpha}}
\newcommand{\sigmasrel}{\sigma\spec{\alpha,\mathrm{rel}}}
\newcommand{\sigmascorr}{\sigma\spec{\alpha,\mathrm{corr}}}
\newcommand{\udots}{\dot{\mathbf u}\spec{\alpha}}
\newcommand{\Tdots}{ \dot{T}\spec{\alpha}}
\newcommand{\sigmadots}{\dot{\sigma}\spec{\alpha}}
\newcommand{\Ps}{p\spec{\alpha}}

\begin{document}
\let\WriteBookmarks\relax
\def\floatpagepagefraction{1}
\def\textpagefraction{.001}
\shorttitle{VHS Multispecies BGK Model}
\shortauthors{M. Pfeiffer et~al.}
\title [mode = title]{A Multispecies ESBGK Model for Gas Mixtures with Variable Hard Sphere Transport: Theory and Verification}    

\author[1]{M. Pfeiffer}
\affiliation[1]{organization={Institute of Space Systems, University of Stuttgart},
                addressline={Pfaffenwaldring 29}, 
                city={Stuttgart},
                postcode={70569}, 
                country={Germany}}

\author[1]{F. Tuttas}

\author[2]{J. Mathiaud}
\affiliation[2]{organization={CNRS, IRMAR, University Rennes},
                city={Rennes},
                postcode={6625}, 
                country={France}}

\author[3]{L. Mieussens}
\affiliation[3]{organization={Univ. Bordeaux, CNRS, Bordeaux INP, IMB, UMR 5251, F-33400 Talence, France},
                addressline={Talence}, 
                city={Bordeaux},
                postcode={5251}, 
                country={France}}


\begin{abstract}
A multi-species Bhatnagar–Gross–Krook (BGK) model for gas mixtures is presented that achieves the correct species-wise relaxation of velocities, temperatures, and pressure tensors according to the Boltzmann collision integral, as well as the correct mixture Prandtl number, while retaining a single relaxation term per species.
The model extends the ellipsoidal statistical BGK (ESBGK) model by introducing relative relaxation targets for each species, derived from the Variable Hard Sphere (VHS) production rates of the Grad 13 approximation.
Three approaches for the species relaxation frequency are proposed and analyzed:
a Grad 13-based per-species frequency, a mixture-averaged frequency, and an empirical harmonic mean of the two.
The model is implemented in the particle-based code PICLas and verified against Direct Simulation Monte Carlo (DSMC) results for a range of test cases, including 0D reservoir relaxation, mass diffusion, supersonic Couette flow, and hypersonic flow around a 70$^\circ$ blunted cone for binary and ternary gas mixtures.
Across all test cases, the proposed model reproduces the correct Prandtl number, species temperature and velocity relaxation rates, and pressure tensor relaxation, with the empirical relaxation frequency consistently yielding the best agreement with DSMC.
\end{abstract}


\maketitle


\section{Introduction}
Significant challenges for numerical methods arise when looking at space applications, micro and nano flows, as well as vacuum technology due to the large density gradients involved, ranging from the continuum to the free molecular flow regime.
Thus, versatile numerical approaches are required to correctly and efficiently handle their multi-scale and non-equilibrium nature.
The well-established Direct Simulation Monte Carlo (DSMC)~\cite{bird} method, even though achieving a highly accurate solution of these flows, requires excessive computational effort in the transition and continuum regimes, which makes a coupling of DSMC with a computationally more efficient method for denser flow regimes desirable.
In the latter case, computational fluid dynamics (CFD) is typically used.
However, coupling with the DSMC method is extremely challenging due to the very different underlying approaches of both methods, with problems especially at the boundaries between DSMC and CFD with the statistical noise of the DSMC method~\cite{zhang2019particle,burt-boyd-ld,pfeiffer2019evaluation,SCHWARTZENTRUBER2006402}.
As a promising alternative, particle-based continuum methods have emerged for a simple and straightforward coupling with DSMC.
Recent developments include but are not limited to the Fokker--Planck approach~\cite{hossein,fp,mathiaud2016fokker,Jun2019} and the Bhatnagar--Gross--Krook (BGK) model~\cite{bgk,gallis-torczynski-2,zhang2019particle,burt2006evaluation}.

The focus of this paper lies on the BGK method, which in general approximates the collision integral of the Boltzmann equation by a relaxation process.
This has the advantage that the mean free path and the collision frequency do not need to be resolved as within the DSMC method, leading to a less restrictive choice of time step and particle weighting factor for particle simulations.
Instead, the time step is determined by the stiffness of the BGK collision term, and thus only the relaxation frequency is to be resolved.
Therefore, the computational efficiency is expected to be higher compared to DSMC simulations, particularly for low-$Kn$ regimes.
Another advantage over DSMC is the possible utilisation of different time integration and space interpolation methods to achieve significantly coarser resolutions.
For a long time, this has been a research focus in the context of discrete velocity methods~\cite{guo2013discrete,mieussens2000discrete}, but recently also progress has been made in the research of particle methods~\cite{PhysRevE.106.025303, fei2021efficient, FEI2020108972, liu2020unified}.

An overview of different BGK models for gas mixtures can be found in~\citet{overview-bgk-mix}.
In general, a distinction is made between two different types of mixture models with multiple relaxation terms on the one hand or only one relaxation term on the other.
The first type includes models~\cite{asinari2008asymptotic,garzo1989kinetic,hamel1965kinetic,klingenberg2018consistent,klingenberg2018kinetic,bobylev2018general,bisi2024mixed}, where for a gas mixture consisting of $N$ species, the right-hand side of the BGK equation then consists of a sum of $N$ relaxation operators per species.
With this, the correct collision rates and thus the energy and moment exchange rates between the individual species can theoretically be reproduced correctly according to the Boltzmann equation.
However, the models become significantly more complex and therefore computationally more expensive.
Also, it is challenging to choose free parameters such as the different relaxation frequencies to achieve e.g. the correct Prandtl number of the gas mixture, in particular when looking at mixtures with more than two components or when additional exchange terms for the internal degrees of freedom are added in the polyatomic case.
The second type of models~\cite{andries2002consistent,brull,brull2021ellipsoidal,todorova-shakhov-esbgk,todorova2020} uses only one relaxation term on the right hand side of the Boltzmann equation.
When looking at the particle-based application of the models, some advantages arise compared to models with multiple relaxation terms:
Despite being less complex and thus less computationally expensive, the collective behavior of the gas mixture, such as the Prandtl number, can be modeled correctly with significantly fewer parameters, whereby established mixing rules can be used.
In addition, solely the moments of the entire mixture, not the individual species, are required, and thus a significantly smaller number of simulation particles can be used compared to multi-relaxation term models, where all individual moments of the distribution function are needed.
However, these models have other disadvantages regarding the relaxation of individual species in a mixture, which is discussed in greater detail at the end of this chapter.

In recent times, the particle-based ellipsoidal statistical BGK (ESBGK)~\cite{esbgk} and the Shakhov BGK~\cite{shakhov} methods have been investigated in particular~\cite{PhysRevE.106.025303, fei2021efficient, FEI2020108972, liu2020unified,piclas-bgk,fei2020benchmark}, since they both produce the correct Prandtl number of the gas.
Different investigations showed that in some cases the ESBGK model is more robust and more efficient in the particle context~\cite{piclas-bgk, pfeiffer2019evaluation}.
Due to this, until now an ESBGK approach based on the models of \citet{MATHIAUD202265,energy-cons} and \citet{brull,brull2021ellipsoidal} was followed, using only one relaxation term.
Most recently, the resulting particle-based ESBGK model was extended to gas mixtures including non-equilibrium states of internal degrees of freedom of diatomic and polyatomic molecules with quantized vibrational states~\cite{bgk-poly,bgk-multispecies,bgk-poly-mix-hild}. Furthermore, a similar SBGK model with the same possibilities was developed~\cite{pfeiffer2026shakhovbasedbhatnagargrosskrookmodelpolyatomic}.

Still, the disadvantage of models with only one relaxation term is that the temperature and velocity relaxation between species cannot or only with constraints be correctly represented.
Additionally, the pressure tensor relaxation is only correct for the entire mixture but not for each individual species, as also discussed in~\cite{frolova2023numerical}.
Thus, in this paper, the model is extended to achieve the correct relaxation rates of these quantities according to the Boltzmann integral in thermal non-equilibrium and at different flow velocities, while still generating the correct relaxation of pressure tensor and heat flux within the limits of the thermal equilibrium of the mixture. The approach is inspired by similar ideas for Fokker-Planck operators applied to mixtures of~\citet{hepp2020kinetic,hepp2020kineticvhs,kim2025particle,kim2025stochastic} and kinetic models~\cite{li2024kinetic,kim2026particle}.

\section{Theory}

The correct exchange rates for species velocity, temperature, and pressure tensor were provided in Gupta's work as production rates for hard sphere gases for the Grad 13 and Grad 26 distributions~\cite{gupta2015mathematical}.
In the model presented in this paper, only the Grad 13 approximations for hard sphere gases are used, considering only the lower-order production term from Gupta.
An extension to VHS was carried out by Hepp~\cite{hepp2020kinetic,hepp2020kineticvhs}, which largely results only in altered prefactors in the production terms, making the transition to VHS straightforward.
Under these assumptions, the following results are obtained for a mixture of $N$ species:
\begin{align} \label{eq:system1}
    \frac{\mathrm{d} \mathbf{u}^{(\alpha)}}{\mathrm{d} t} = \dot{\mathbf u}^{(\alpha)} &=-\sum_{\beta=1}^N \nu^{(\alpha \beta,\mathrm{VHS})} \mu^{(\beta\alpha)}\left[\frac{5}{3}\xi^{\mathrm{VHS},1}\left(\mathbf{u}^{(\alpha)}-\mathbf{u}^{(\beta)}\right) + \frac{\xi^{\mathrm{VHS},2}}{6\Theta^{(\alpha \beta)}}\left(\frac{\mathbf{h}^{(\alpha)}}{\rho^{(\alpha)}}-\frac{\mathbf{h}^{(\beta)}}{\rho^{(\beta)}}\right)\right],\\
    \frac{\mathrm{d} T^{(\alpha)}}{\mathrm{d} t} = \dot{T}^{(\alpha)} &= -\frac{10}{3k_\mathrm{B} n^{(\alpha)}} \sum_{\beta=1}^N \nu^{(\alpha \beta,\mathrm{VHS})} \mu^{(\beta\alpha)}\rho^{(\alpha)}\xi^{\mathrm{VHS},1}\left[\Theta^{(\alpha\beta)}\Delta \Theta^{(\alpha\beta)}- \frac{1}{3}\left(\mu^{(\alpha\beta)}-\mu^{(\beta\alpha)}\right)u_{d,i}^{(\alpha)}u_{d,i}^{(\beta)}\right],\label{eq:system2}\\
    \frac{\mathrm{d} \sigma^{(\alpha)}_{ij}}{\mathrm{d} t} = \dot{\sigma}^{(\alpha)}_{ij} &= -\sum_{\beta=1}^N \nu^{(\alpha \beta, \mathrm{VHS})} \mu^{(\beta\alpha)}\left[4\mu^{(\beta\alpha)}\left\{\sigma^{(\alpha)}_{ij}+\frac{\xi^{\mathrm{VHS},3}}{3}\left(\sigma^{(\alpha)}_{ij}-\frac{\rho^{(\alpha)}}{\rho^{(\beta)}}\sigma^{(\beta)}_{ij}\right)\right\}\right.\\
    &\left.+\frac{10}{3}\xi^{\mathrm{VHS},1}\left(\mu^{(\alpha\beta)}-\mu^{(\beta\alpha)}\right)\sigma^{(\alpha)}_{ij}\right].
    \label{eq:system3}
\end{align}
The following variables are used in the equations for species $\alpha$ and $\beta$:
\begin{alignat}{1}
    &\mu\spec{\alpha\beta} =\frac{m\spec{\alpha}}{m\spec{\alpha}+m\spec{\beta}},\\
    &\Theta\spec{\alpha} = \frac{k_\mathrm{B} T\spec{\alpha}}{m\spec{\alpha}},\\
    &\Theta\spec{\alpha\beta} = \frac{\Theta\spec{\alpha}+\Theta\spec{\beta}}{2},\\
    &\Delta\Theta\spec{\alpha\beta} =\frac{\mu\spec{\alpha\beta}\Theta\spec{\alpha}-\mu\spec{\beta\alpha}\Theta\spec{\beta}}{\Theta\spec{\alpha\beta}},\\
    &\mathbf{h}\spec{\alpha}=\mathbf{q}\spec{\alpha}-\frac{5}{2}\rho\spec{\alpha}\Theta\spec{\alpha}\mathbf{u}_\mathrm{d}\spec{\alpha},\\
    &\nu\spec{\alpha\beta,\mathrm{VHS}}=\frac{16}{5}\xi^{\mathrm{VHS},4}n\spec{\beta}\sqrt{\pi\Theta\spec{\alpha\beta}}\left(\frac{d\spec{\alpha}+d\spec{\beta}}{2}\right)^2.
\end{alignat}
Here, $n\spec{\alpha}$ is the particle density of species $\alpha$, $\rho\spec{\alpha}$ is the mass density with mass $m\spec{\alpha}$, $T\spec{\alpha}$ is the temperature, $\nu\spec{\alpha\beta,\mathrm{VHS}}$ is the VHS collision frequency with the particle diameter $d\spec{\alpha}$, $\mathbf{q}\spec{\alpha}$ is the heat flux and $\mathbf{u}_\mathrm{d}\spec{\alpha}$ is the relative flow velocity of species $\alpha$ with respect to the bulk flow velocity of the mixture $\mathbf u$.
In addition, the pressure tensor $p_{ij}\spec{\alpha}$ as well as the trace free pressure tensor $\sigma_{ij}\spec{\alpha}$ can be defined with the thermal particle velocity $\mathbf c=\mathbf v -\mathbf u$ from the particle velocity $\mathbf v$, the total energy $E$ and the thermal energy $\mathcal{E}$:
\begin{alignat}{1}
	&n\spec{\alpha}=\int_{\mathbb{R}^3} f\spec{\alpha}\,\mathrm d\mathbf v, \quad n\spec{\alpha}\mathbf u\spec{\alpha}=\int_{\mathbb{R}^3} \mathbf v f\spec{\alpha}\,\mathrm d\mathbf v, \\
	&\mathcal{E}\spec{\alpha}=\frac{3}{2}k_{\mathrm{B}} T\spec{\alpha}= \frac{m\spec{\alpha}}{2n\spec{\alpha}}\int_{\mathbb{R}^3} \left|\mathbf{c}\right|^2 f\spec{\alpha}\,\mathrm d\mathbf v,\\
	&E\spec{\alpha}=\frac{1}{2}m\spec{\alpha}n\spec{\alpha}\mathbf u^{(\alpha)2} + n\spec{\alpha}\mathcal{E}\spec{\alpha},\\
    &\mathbf q\spec{\alpha}=m\spec{\alpha}\int_{\mathbb{R}^3}  \mathbf c \left|\mathbf{c}\right|^2 f\spec{\alpha}\,\mathrm d\mathbf v,\\
    &p_{ij}\spec{\alpha}=m\spec{\alpha}\int_{\mathbb{R}^3}  c_ic_j f\spec{\alpha}\,\mathrm d\mathbf v,\\
    &\sigma\spec{\alpha}_{ij}=m\spec{\alpha}\int_{\mathbb{R}^3}  c_{\langle i}c_{j\rangle}f\spec{\alpha}\,\mathrm d\mathbf v.
\end{alignat}
Furthermore, the macroscopic mean values of the flow are given by:
\begin{alignat}{1}
	&\rho=\sum_{\alpha=1}^N m\spec{\alpha} n\spec{\alpha},\quad \rho \mathbf u = \sum_{s=1}^N m\spec{\alpha} n\spec{\alpha} \mathbf u\spec{\alpha},\\    
	&n\mathcal{E} + \frac{\rho}{2}\mathbf u^2 = E=\sum_{\alpha=1}^N E\spec{\alpha},
	\quad \mathcal{E}=\frac{3}{2}k_{\mathrm{B}} T,\\
    &\mathbf u_\mathrm{d}\spec{\alpha} = \mathbf u\spec{\alpha} - \mathbf u,\\
    &p_{ij}=\sum_{\alpha=1}^N p_{ij}\spec{\alpha}.
\end{alignat}
Moreover, the VHS factors $\xi$ are given in~\cite{hepp2020kineticvhs}:
\begin{alignat}{1}
	&\xi^{\mathrm{VHS},1}=3\frac{\Gamma(3-\hat \omega\spec{\alpha\beta})}{\Gamma(4-\hat \omega\spec{\alpha\beta})},\\
    &\xi^{\mathrm{VHS},2}=6-15\frac{\Gamma(3-\hat \omega\spec{\alpha\beta})}{\Gamma(4-\hat \omega\spec{\alpha\beta})},\\
    &\xi^{\mathrm{VHS},3}=\frac{15}{2}\frac{\Gamma(3-\hat \omega\spec{\alpha\beta})}{\Gamma(4-\hat \omega\spec{\alpha\beta})}-\frac{3}{2},\\
    &\xi^{\mathrm{VHS},4}=\frac{1}{6}c_{\mathrm{ref}}^{(\alpha\beta)2\hat \omega\spec{\alpha\beta}}\Gamma(4-\hat \omega\spec{\alpha\beta})\left(2\Theta\spec{\alpha}+2\Theta\spec{\beta}\right)^{-\hat \omega\spec{\alpha\beta}},\\
\end{alignat}
with 
\begin{alignat}{1}
    &\hat \omega\spec{\alpha\beta} = \omega\spec{\alpha\beta,\mathrm{VHS}}-0.5,\\
    &c_{\mathrm{ref}}^{(\alpha\beta)2\hat \omega\spec{\alpha\beta}}=\frac{1}{\Gamma\left(2-\hat \omega\spec{\alpha\beta}\right)}\left(\frac{2k_BT_{\mathrm{ref}}\spec{\alpha\beta}}{m_\mathrm{r}\spec{\alpha\beta}}\right)^{\hat \omega\spec{\alpha\beta}},\\
\end{alignat}
using the VHS parameter $\omega\spec{\alpha\beta,\mathrm{VHS}}$, the VHS reference temperature $T_{\mathrm{ref}}\spec{\alpha\beta}$, and the reduced mass
\begin{equation}
    m_\mathrm{r}\spec{\alpha\beta}=\frac{m\spec{\alpha}m\spec{\beta}}{m\spec{\alpha}+m\spec{\beta}}.
\end{equation}
The second term in Eq.~\eqref{eq:system1} is responsible for thermal diffusion. 

\paragraph{Proposed ESBGK model}
Using these equations in the first step, the proposed ESBGK model is as follows
\begin{alignat}{1}
    &\partial_t f\spec{\alpha} + \mathbf {v} \partial_{\mathbf x} f\spec{\alpha} = \nu\spec{\alpha}(f^{(\alpha)t}-f\spec{\alpha}), \label{eq:BGK}\\
   &f^{(\alpha)t}=\frac{n\spec{\alpha}}{\sqrt{\det 2\pi\mathcal A\spec{\alpha}}} \exp\left[-\frac{1}{2}(\mathbf v-\mathbf u\spec{\alpha,\mathrm{rel}})^\mathrm{T} \mathcal {A\spec{\alpha}}^{-1} (\mathbf v-\mathbf{u}\spec{\alpha,\mathrm{rel}})\right], \label{eq:BGK2} \\
   &\mathcal {A\spec{\alpha}}= \frac{k_\mathrm{B} T\spec{\alpha,\mathrm{rel}}}{m\spec{\alpha}} \mathcal I + \frac{1}{n\spec{\alpha}m\spec{\alpha}}\sigma\spec{\alpha,\mathrm{rel}}_{ij}. \label{eq:BGK3}
\end{alignat}
The matrix $\mathcal A$ is used to produce the correct relaxation of the pressure tensor for each species.
To fulfill Eqs.~\eqref{eq:system1}, \eqref{eq:system2} and \eqref{eq:system3}, the idea of relative relaxation is used again from diatomic and polyatomic molecules~\cite{MATHIAUD202265,bgk-poly-mix-hild} and a relative temperature $T\spec{\alpha,\mathrm{rel}}$, a relative flow velocity $\mathbf u\spec{\alpha,\mathrm{rel}}$, and a relative traceless pressure tensor $\sigma \spec{\alpha,\mathrm{rel}}_{ij}$ per species are introduced:
\begin{alignat}{1}
    &\mathbf u\spec{\alpha,\mathrm{rel}}=\mathbf u\spec{\alpha} + \frac{1}{\nu\spec{\alpha}}\dot{\mathbf u}\spec{\alpha} \label{eq:rate_u},\\
    &T\spec{\alpha,\mathrm{rel}} = T\spec{\alpha} + \frac{1}{\nu\spec{\alpha}}\dot{T}\spec{\alpha}- T\spec{\alpha,\mathrm{corr}} \label{eq:rate_T},\\
    &\sigma\spec{\alpha,\mathrm{rel}}_{ij} = \sigma\spec{\alpha}_{ij} + \frac{1}{\nu\spec{\alpha}} \dot{\sigma}\spec{\alpha}_{ij}- \sigma\spec{\alpha,\mathrm{corr}}_{ij}. \label{eq:rate_sigma}
\end{alignat}    
Additional correction terms $T\spec{\alpha,\mathrm{corr}}$ and $\sigma\spec{\alpha,\mathrm{corr}}_{ij}$ appear here, which are created by the BGK method itself.
The problem is that the target distribution function has a different species velocity $\mathbf u\spec{\alpha,\mathrm{rel}}$ than the bulk velocity $\mathbf u$.
This results in a relative velocity $\mathbf{\hat u}\spec{\alpha,\mathrm{rel}} = \mathbf u\spec{\alpha,\mathrm{rel}} - \mathbf u$ with respect to the bulk velocity of the mixture.
Energetically, these are parts of the resulting temperature, as well as pressure tensor:
\begin{alignat}{1}
    T\spec{\alpha,\mathrm{corr}}&= \frac{m\spec{\alpha}}{3k_\mathrm{B}}\left(\mathbf {\hat u}\spec{\alpha,\mathrm{rel}}\right)^2,\\
    \sigma\spec{\alpha,\mathrm{corr}}_{ij}&=m\spec{\alpha}n\spec{\alpha}\hat u\spec{\alpha,\mathrm{rel}}_{\langle i}\hat u\spec{\alpha,rel}_{j\rangle}.
\end{alignat}

\paragraph{Definition of the relaxation frequency}\label{sec:relaxfreqdef}
In the proposed ESBGK model, $\nu\spec{\alpha}$ is the relaxation frequency of each species used to fix the relaxation of the heat flux of the mixture and produce the correct Prandtl number as usual for ESBGK models.
The relaxation frequency can be different for each species.

Different approaches can be used to determine the relaxation frequency.
The first approach defines a relaxation frequency per species based on the temporal evolution of the heat flux vector in the Grad 13 approximation~\cite{gupta2015mathematical}.
The drawback of this BGK formulation is that the heat flux vector can only decrease in magnitude, whereas, according to the Grad 13 equation, it may also grow under strongly non-equilibrium conditions through the exchange with species that have very different heat flux values.
To capture this behavior, one would have to replace the ESBGK target function with a different target function capable of generating heat flux.
As an example, an extension of the ES-Fokker-Planck method was proposed in~\cite{kim2025particle} to reproduce the heat flux correctly.

Moreover, in the Grad 13 equation each component of the heat flux vector can, in principle, relax at a different rate depending on the degree of non-equilibrium, which clearly cannot be modeled with a single relaxation frequency.
To remain within the ESBGK framework, the idea is therefore to retain only the first dominant term of the heat flux relaxation from the Grad 13 distribution and to neglect the higher-order non-equilibrium exchange terms, such as those arising from a heat flux difference between individual species.
This has the additional advantage that the first dominant term is identical for all components of the heat flux vector, so it can be captured easily by a single relaxation frequency per species.
A further assumption is that the relevant heat flux $\mathbf q\spec{\alpha}$ is larger than $m\spec{\alpha}n\spec{\alpha}\Theta\spec{\alpha}\mathbf{u}_d\spec{\alpha}$;
otherwise an additional correction term would arise that is different component-wise and could therefore not be represented by a single relaxation frequency. Under these assumptions, the relaxation frequency per species reads:
\begin{equation}
    \nu\spec{\alpha}_\mathrm{Grad13}= \sum_{\beta=1}^N 4 Pr\spec{\alpha}\nu^{(\alpha \beta,\mathrm{VHS})}\mu\spec{\beta\alpha}\mu\spec{\beta\alpha}.\label{eq:nu-grad13}
\end{equation}
Here, $Pr\spec{\alpha}$ is the Prandtl number of a species, which equals $Pr\spec{\alpha}=\nicefrac{2}{3}$ for monoatomic species.
Remarkably, despite the assumptions outlined above, the proposed model yields very good results even under relatively strong non-equilibrium conditions, as demonstrated in the results section.

A second way to define the relaxation frequency is to use the average relaxation frequency of the mixture for all species, as shown in~\citet{bgk-multispecies,bgk-poly-mix-hild}.
This yields the correct Prandtl number for the mixture and, with the use of collision integrals, can also achieve very high accuracy, as demonstrated in~\cite{bgk-multispecies,bgk-poly-mix-hild}.
The resulting relaxation frequency for the mixture depends on the thermal conductivity $K$ and the specific heat $c_{\mathrm{p}}$, or alternatively on the mixture viscosity $\mu_\mathrm{vis}$ and the mean mixture Prandtl number $Pr_\mathrm{mean}$:
\begin{equation}
    \nu\spec{\alpha}_\mathrm{mean}=\nu_\mathrm{mean} = \frac{nk_{\mathrm{B}} T}{K}\gamma c_{\mathrm{p}}=\frac{nk_{\mathrm{B}} T}{\mu_\mathrm{vis}}\gamma Pr_\mathrm{mean}.\label{eq:nu-mean}
\end{equation}
Here, $\gamma$ is the Prandtl correction factor introduced in~\cite{brull,bgk-multispecies,bgk-poly-mix-hild}.
There is, however, a fundamental difference between the derivation of the Prandtl number and its correction factor in these references and the approach adopted in this work.
The assumption made in~\citet{brull,bgk-multispecies,bgk-poly-mix-hild} that all species share the same temperature $T^{(\alpha)} = T$ no longer holds.
Instead, in the proposed model the Prandtl number depends not only on the densities $\rho^{(\alpha)}$ and masses $m^{(\alpha)}$, but also on the individual temperatures $T^{(\alpha)}$ and, strictly speaking, on their temperature gradients $\nabla T^{(\alpha)}$.
To avoid increasing the complexity of the model further, we assume that $\nabla T^{(\alpha)} = \nabla T$.
Although this may appear to be a strong simplification, the results section will show that good accuracy can still be obtained for estimating $\nu\spec{\alpha}_{\mathrm{mean}}$ at a manageable level of complexity.
Under these assumptions, the Prandtl correction factor is given by:
\begin{align}
	&\gamma = \frac{m}{n} \frac{\sum_{s=1}^M \frac{n_s}{m_s}T^{(\alpha)} }{T},\\
	&m=\sum_{s=1}^M \frac{n_s}{n}m_s,
	\quad n=\sum_{s=1}^M n_s. \label{eq:m}
\end{align}
In addition,~\citet{bgk-multispecies,bgk-poly-mix-hild} describe in detail how $K$ and $c_{\mathrm{p}}$ can be computed for the mixture using collision integrals.

The results in Section~\ref{sec:ver} show that the average relaxation frequency $\nu\spec{\alpha}_\mathrm{mean}$ produces good heat flux relaxation rates for the mixture as a whole, but over- or underestimates those of the individual species.
This is to be expected, since a single relaxation frequency is used for all species.
The results further show that the per-species relaxation frequency of the Grad 13 approximation $\nu\spec{\alpha}_\mathrm{Grad13}$ behaves in exactly the opposite manner:
Wherever the mean relaxation frequency $\nu\spec{\alpha}_\mathrm{mean}$ overestimates the species heat flux relaxation, the Grad 13 relaxation frequency $\nu\spec{\alpha}_\mathrm{Grad13}$ underestimates it, and vice versa.
This is most likely a consequence of the fact, explained above, that in an ESBGK model only the dominant first term of the Grad 13 approximation can be used for relaxation.
For this reason, a third, empirical approach is proposed here, in which the harmonic mean of the Grad 13 and the mean relaxation frequencies is used:
\begin{equation}
    \nu\spec{\alpha}_\mathrm{empi}=\frac{2}{(1/\nu_\mathrm{mean}+ 1/\nu\spec{\alpha}_\mathrm{Grad13})}.\label{eq:nu-empi}
\end{equation}
This yields the correct arithmetic mean of the relaxation times of the two approaches.
As shown by the results in Section~\ref{sec:ver}, the empirical relaxation frequency matches both the mean heat fluxes and the heat flux relaxation frequencies of the individual species relatively well.


\subsection{Limits of the Model}\label{sec:limitsmet}
The proposed model with its three different options for the relaxation times comes with several theoretical limitations.
The most obvious one is that in Eq.~\eqref{eq:rate_T} the species temperature $T^{(\alpha,\mathrm{rel})}$ may, in principle, become negative for certain species.
This can occur either due to the change term $\dot{T^{(\alpha)}}$ in cases of extreme non-equilibrium with very large mass disparities, or due to the subtraction of the correction term $T^{(\alpha,\mathrm{corr})}$.
The latter may become very large when the species velocities differ significantly from the bulk velocity, e.g. in situations where species stream in different directions.
Both situations are very uncommon in simulations, since such strong non-equilibrium states are rarely produced.
In the verification section~\ref{sec:ver}, for instance in the mass-diffusion examples, cases are shown where species move in different directions and are still handled successfully by the model. 

If the issue occurs nevertheless, we recommend the following corrective procedure.
As a first step, set 
\begin{equation}
    \mathbf{u}^{(\alpha,\mathrm{rel})} = \mathbf{u}
\end{equation}
for each species, which immediately implies $T^{(\alpha,\mathrm{corr})} = 0$.
This introduces an error only in the relaxation of the species velocities, while momentum conservation is still preserved.
If after this adjustment the value of $T^{(\alpha,\mathrm{rel})}$ remains negative, the next step is to set 
\begin{equation}
    T^{(\alpha,\mathrm{rel})} = T
\end{equation}
for all species.
This introduces an additional error in the temperature relaxation for that cell, but energy conservation is still guaranteed. 

A less obvious issue can arise when the relaxation frequency $\nu^{(\alpha)}_{\mathrm{Grad13}}$ is chosen.
In cases of extreme mass and density ratios, the relaxation frequency of certain species may become very small because their contribution to the heat flux change is minor.
However, in \eqref{eq:rate_u},\eqref{eq:rate_T},\eqref{eq:rate_sigma}, this may cause the relaxation terms $\mathbf{u}^{(\alpha,\mathrm{rel})}$, $T^{(\alpha,\mathrm{rel})}$, and $\sigma^{(\alpha,\mathrm{rel})}_{ij}$ to become excessively large in order to compensate for such small frequencies. 
As discussed in one of the mass-diffusion examples in the verification section~\ref{sec:massdiff}, this can lead to numerical difficulties.

\subsection{Equilibrium}
We verify that if the collision operator of Eq.~\eqref{eq:BGK} vanishes for every species, i.e.
\begin{equation} \label{eq:equilibrium}
    \fs = \fst \quad \text{for all } \alpha,
\end{equation}
then
\begin{equation}
    \fs = M[\rhos, u, T] \quad \text{for every } \alpha.
\end{equation}

\paragraph{Step 1: Vanishing of exchange rates.}
We first show that all exchange rates defined in Eqs.~\eqref{eq:rate_u}--\eqref{eq:rate_sigma} are zero.

\vspace{2mm}
\emph{First-order moment.} Condition~\eqref{eq:equilibrium} directly gives $\us = \usrel$, and substituting into Eq.~\eqref{eq:rate_u} yields $\udots = 0$.

\vspace{2mm}
\emph{Second-order moment.} Taking the tensorial second-order moment of Eq.~\eqref{eq:equilibrium} gives
\begin{equation} \label{eq:peq}
    \Ps = \rhos \mathcal{A}^{(\alpha)} + \rhos (\us - u) \otimes (\us - u).
\end{equation}
Taking the trace, and using the definitions of $\Tsrel$ and $\uhatsrel$, we obtain $\Tdots = 0$.
Inserting this into Eq.~\eqref{eq:peq}, together with the decomposition
\begin{equation}
    \Ps = \sigmas + k_\mathrm{B} \ns \Ts \mathcal{I}
\end{equation}
and the definitions of $\sigmasrel$ and $\sigmascorr$, then yields $\sigmadots = 0$.
Finally, substituting $\sigmadots = 0$ back into Eq.~\eqref{eq:peq} gives $\rhos\,|\us - u|^2 = 0$, and hence $\us = u$ for all $\alpha$.

\paragraph{Step 2: Equality of temperatures.}
Using the result $\Tdots = 0$ in the definition of $\dot{T}^{(\alpha)}$ from Eq.~\ref{eq:system2}, one finds
\begin{equation}
    \Ts = \sum_{\beta} w^{(\alpha\beta)}\, T^{(\beta)},
\end{equation}
where $w^{(\alpha\beta)} \in (0,1)$ and $\sum_\beta w^{(\alpha\beta)} = 1$ for all $\alpha$.
Each temperature $\Ts$ is thus a strict convex combination of all species temperatures, which is only possible if all temperatures are equal.
Using the definition of the mixture temperature $T$ then gives $\Ts = T$ for all $\alpha$. This argument follows~\cite{andries2002consistent}.

\paragraph{Step 3: Vanishing of stress deviators.}
With the results of Steps~1 and~2, Eq.~\eqref{eq:peq} reduces to $\rhos \mathcal{A}^{(\alpha)} = \Ps$.
Since $\fs = G[\ns, u, \Ps/\rhos]$ and $\Ts = T$, proving $\fs = M[\ns, u, T]$ is equivalent to showing
\begin{equation}
    \Ps = \ns k_\mathrm{B} T \mathcal{I} \quad \text{i.e., } \sigmas = 0.
\end{equation}
The convexity argument of Step~2 cannot be applied here.
Instead, we construct a quadratic form.
For any indices $i$ and $j$, the right-hand side of Eq.~\eqref{eq:system3} is multiplied by $\sigma^{(\alpha)}_{ij}/(\ns)^2$ and summed over $\alpha$.
Let $L$ denote the result.
We aim to show that $L > 0$ whenever at least one $\sigma^{(\alpha)}_{ij}$ is not zero; since $L = 0$ (because $\sigmadots = 0$), this forces all $\sigma^{(\alpha)}_{ij} = 0$.

Separating the double sum into diagonal and symmetrized off-diagonal parts, one obtains
\begin{equation}
\begin{split}
   L = {} & \sum_{\alpha,\beta}
     \nu^{(\alpha\beta,\mathrm{VHS})} \mu^{(\beta\alpha)}
     \left[\!\left(4 - \tfrac{10}{3}\xi^{\mathrm{VHS},1}\right)\mu^{(\beta\alpha)}
     + \tfrac{10}{3}\xi^{\mathrm{VHS},1}\mu^{(\alpha\beta)}\right]
     \frac{(\sigma^{(\alpha)}_{ij})^2}{(\ns)^2} \\
   & + \frac{1}{2}\sum_{\alpha,\beta}
     \nu^{(\alpha\beta,\mathrm{VHS})}\,\frac{4\xi^{\mathrm{VHS},3}}{3}
     \left(\frac{m^{(\alpha)} m^{(\beta)}}{m^{(\alpha)} + m^{(\beta)}}\right)^{\!2}
     \left(\frac{\sigma^{(\alpha)}_{ij}}{\rho^{(\alpha)}}
          + \frac{\sigma^{(\beta)}_{ij}}{\rho^{(\beta)}}\right)^{\!2}.
\end{split}
\end{equation}
The second sum is manifestly nonnegative and vanishes only if all $\sigma^{(\beta)}_{ij} = 0$.
The first sum is nonnegative under the condition $4 - \tfrac{10}{3}\xi^{\mathrm{VHS},1} > 0$, which is equivalent to $\omega^{(\alpha\beta,\mathrm{VHS})} < 1$ (using the relevant definitions and the identity $\Gamma(z+1) = z\Gamma(z)$).
Under this condition, the first sum also vanishes only if all $\sigma^{(\beta)}_{ij} = 0$.
This concludes the proof: the BGK model admits the correct equilibrium whenever the collision operator vanishes for every species.

The condition $\omega^{(\alpha\beta,\mathrm{VHS})} < 1$ is sufficient for the above proof and appears necessary if one requires the result to hold for arbitrary molecular masses.
Indeed, if the condition fails, one can find a sufficiently large $m^{(\beta)}$ such that $L < 0$.
In practice, when $\omega^{(\alpha\beta,\mathrm{VHS})}$ is defined as the average of $\omega^{(\alpha,\mathrm{VHS})}$ and $\omega^{(\beta,\mathrm{VHS})}$, the condition is satisfied for all species with $\omega^{(\alpha,\mathrm{VHS})} < 1$ (see~\cite{bird}).
However, some species listed in~\cite{bird,pfeiffer2022optimized} do not satisfy this bound.
For these cases, the above proof cannot be used for arbitrary masses.

\section{Implementation}
The proposed method is implemented in PICLas~\cite{piclas} using a particle-based approach.
However, the model can also be used in deterministic simulations, such as in the context of Discrete Velocity methods.

The underlying concept of the particle method follows the implementation presented~\citet{bgk-multispecies,bgk-poly-mix-hild} for the old ESBGK model.
The main distinction between the two implementations lies in the additional computation of the relaxation frequency $\nu\spec{\alpha}$ (according to Eqs.~\ref{eq:nu-grad13}, \ref{eq:nu-mean}, \ref{eq:nu-empi}), relative temperature $T\spec{\alpha,\text{rel}}$, relative flow velocity $\mathbf{u}\spec{\alpha,\text{rel}}$, and relative trace less pressure tensor $\sigma\spec{\alpha,\text{rel}}_{ij}$ (according to Eqs.~\ref{eq:rate_T}, \ref{eq:rate_u}, \ref{eq:rate_sigma}) as well as the adaption of the relaxation term per species as described in Eqs.~\ref{eq:BGK}, \ref{eq:BGK2}, \ref{eq:BGK3}.
Since the primary focus of the present work is the presentation of the new BGK model rather than the particle method itself, details concerning the computation of internal energies, the enforcement of energy and momentum conservation, and the general framework of the stochastic particle approach are not repeated here and can be found in Ref.~\cite{bgk-multispecies,bgk-poly-mix-hild}.

\section{Verification}
\label{sec:ver}
Reservoir simulations were performed for verification purposes, including 0D relaxation processes in a reservoir, mass diffusion test cases, Couette flows, and different flows around a 70$^\circ$ blunted cone.
In the following, the results are compared to the ones of the well-established DSMC method and the previously used ESBGK mixture model~\cite{bgk-multispecies}.
The species used are defined in Table~\ref{tab:specres}.
\begin{table}[ht]
\centering
\begin{tabular}{|c|c|c|c|c|}
\hline
\textbf{Species} & \({d_{\text{ref}}}\) / m& \({T_{\text{ref}}}\) / K& \({\omega_{\text{ref}}}\) & m / kg\\ 
\hline
Argon Ar & $4.05\cdot 10^{-10}$ & 273 &0.77 & $6.6\cdot 10^{-26}$ \\ 
\hline
Helium He &$2.33\cdot 10^{-10}$ & 273 &0.77 & $6.65\cdot 10^{-27}$  \\ 
\hline
Nitrogen N & $3\cdot 10^{-10}$ & 273 &0.74 & $2.36\cdot 10^{-26}$  \\ 
\hline
Oxygen O & $3\cdot 10^{-10}$ & 273 &0.74 & $2.66\cdot 10^{-26}$  \\ 
\hline
\end{tabular}
\caption{Table with reference values \(d_{\text{ref}}\), \(T_{\text{ref}}\), and \(\omega_{\text{ref}}\) for different species.\label{tab:specres}}
\end{table}

In general, the different results are labeled as follows:
While \textit{ESBGK} marks the previous ESBGK model~\cite{bgk-multispecies}, \textit{ESBGK-Grad13}, \textit{ESBGK-mean}, and \textit{ESBGK-empi} denote the newly proposed ESBGK model using the different versions of the relaxation frequency presented in Section~\ref{sec:relaxfreqdef}.
Whenever only the \textit{ESBGK-Grad13} result is shown in the plots, the results of the different relaxation frequencies $\nu$ are identical.
Otherwise, the different results are plotted.

\subsection{0D Reservoir Simulations}
Various species with different initial states were defined in an adiabatic box and then allowed to relax into a thermal equilibrium.

\subsubsection{Case 1}
Case 1 is the simplest scenario, intended to verify whether the correct Prandtl number can be produced in the isothermal case, as is the case in the previously used ESBGK model with only one relaxation term.
To represent different mass ratios, Ar and N are initialized in the first instance, while Ar and He are used in the second instance, with the parameters listed in Table~\ref{tab:initcase1}.
The gas is initialized with the Grad 13 distribution function $f^\text{Grad13}$ to generate a heat flux and a pressure tensor at the beginning of the simulation
\begin{equation}
f^\text{Grad13} =f^\text{M}\left[1+\frac{m^2p_{\langle ij \rangle}}{2\rho k_\text{B}^2 T^2}c_{\langle i}c_{j\rangle}-\frac{m^2q_j c_j}{\rho k_\text{B}^2 T^2}\left(1-\frac{m}{5 k_\text{B}T}\mathbf c^2\right)\right].
\label{eq:grad13}
\end{equation}
\begin{table}[ht]
\centering
\begin{tabular}{|c|c|c|c|c|}
\hline
&\textbf{Species} & \({n_{\text{init}}}\) / m$^{-3}$ & \({T_{\text{init}}}\) / K& \({\mathbf u_{\text{init}}}\) / \si{\meter\per\second} \\ 
\hline
\textbf{Instance 1} &Argon Ar & $2\cdot 10^{22}$ & 5000 & $(0,0,0)^\text{T}$ \\ 
\hline
&Nitrogen N & $6\cdot 10^{21}$ & 5000 & $(0,0,0)^\text{T}$ \\ 
\hline
\textbf{Instance 2} &Argon Ar & $2\cdot 10^{22}$ & 5000 & $(0,0,0)^\text{T}$ \\ 
\hline
&Helium He & $6\cdot 10^{21}$ & 5000 & $(0,0,0)^\text{T}$ \\ 
\hline
\end{tabular}
\caption{Initial parameters for 0D reservoir simulation, case 1.\label{tab:initcase1}}
\end{table}

\begin{figure}[!t]\centering
  \subfloat[Pressure tensor]{\includegraphics[width=0.35\linewidth]{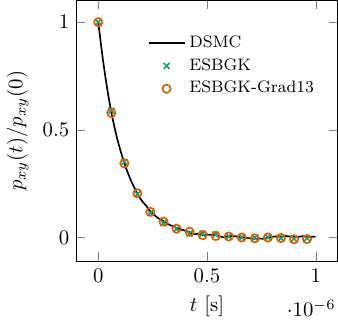}}\quad
  \subfloat[Heat flux]{\includegraphics[width=0.35\linewidth]{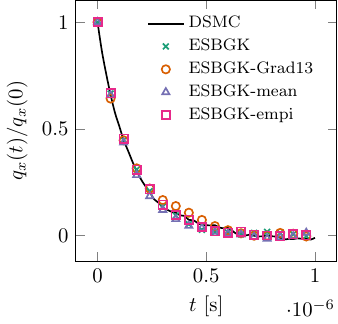}}\\
  \subfloat[Pressure tensor Ar]{\includegraphics[width=0.35\linewidth]{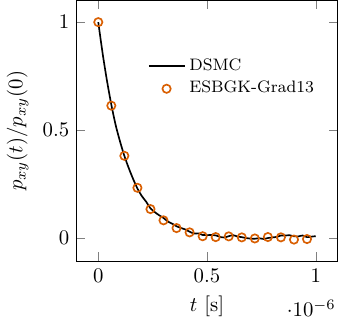}}\quad
  \subfloat[Pressure tensor N]{\includegraphics[width=0.35\linewidth]{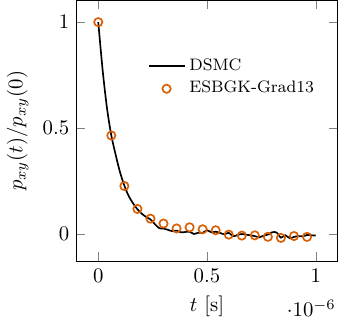}}\\
  \subfloat[Heat flux Ar]{\includegraphics[width=0.35\linewidth]{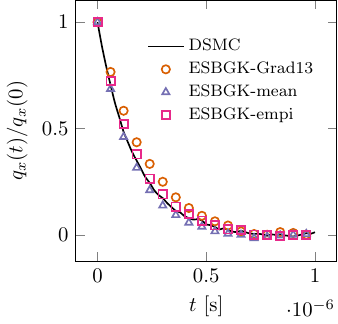}}\quad
  \subfloat[Heat flux N]{\includegraphics[width=0.35\linewidth]{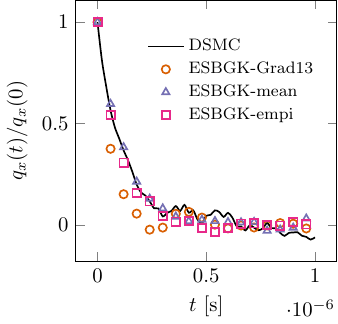}}
  \caption{Pressure tensor and heat flux relaxation with DSMC, ESBGK and the newly proposed ESBGK mixture model 0D reservoir simulation, case 1 with Ar and N.}\label{fig:homreledspbgklin}
\end{figure}

\begin{figure}[!t]\centering
  \subfloat[Pressure tensor]{\includegraphics[width=0.35\linewidth]{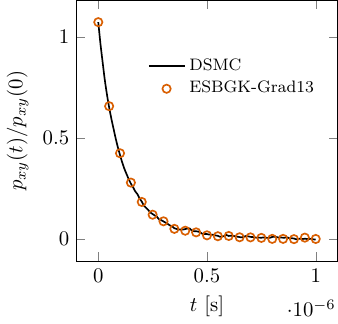}}\quad
  \subfloat[Heat flux]{\includegraphics[width=0.35\linewidth]{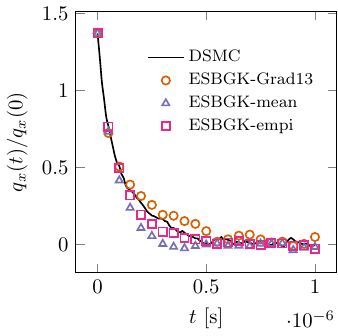}}\\
  \subfloat[Pressure tensor Ar]{\includegraphics[width=0.35\linewidth]{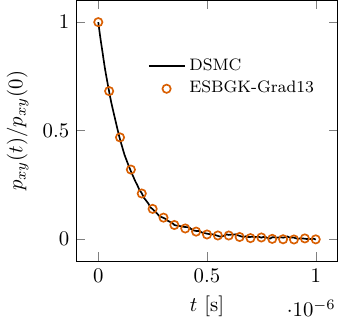}}\quad
  \subfloat[Pressure tensor He]{\includegraphics[width=0.35\linewidth]{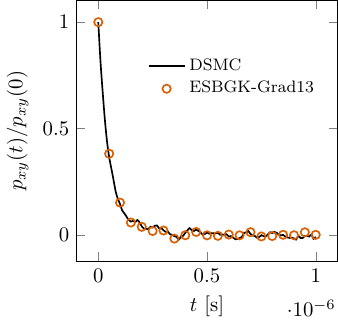}}\\
  \subfloat[Heat flux Ar]{\includegraphics[width=0.35\linewidth]{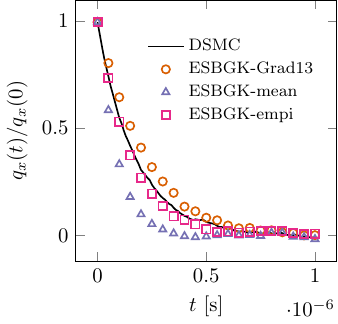}}\quad
  \subfloat[Heat flux He]{\includegraphics[width=0.35\linewidth]{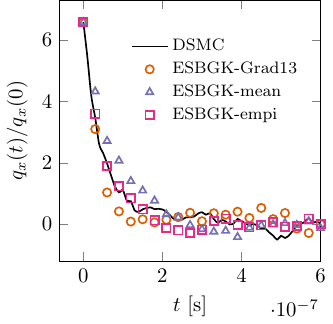}}
  \caption{Pressure tensor and heat flux relaxation with DSMC, ESBGK and the newly proposed ESBGK mixture model 0D reservoir simulation, case 1 with Ar and He.}\label{fig:homreledspbgklinArHe}
\end{figure}

The relaxation for instance 1 is shown in Fig.~\ref{fig:homreledspbgklin}.
A very good agreement between the DSMC solution and the previous ESBGK model as well as the newly presented ESBGK model can be seen for the mixture pressure tensor and the heat flux.
Additionally, the pressure tensor relaxation of the individual species is very good for all three relaxation frequencies. 
However, slight differences between the various relaxation frequencies can be observed in the heat flux results of the individual species. 
As already mentioned in Section~\ref{sec:relaxfreqdef}, \textit{ESBGK-Grad13} and \textit{ESBGK-mean} overestimate and underestimate exactly the opposite species, which is why the empirical approach \textit{ESBGK-empi} performs best here.
The proposed ESBGK model also generates the correct Prandtl number for the isothermal case.
With Ar and He, instance 2 results in the same test case but with a significantly larger mass difference between the species, which makes the case considerably more demanding.
The results are shown in Fig.~\ref{fig:homreledspbgklinArHe}. 
The pressure tensor relaxation agrees very well with DSMC and is again identical for all relaxation frequencies, which is to be expected, since the equation is the same for all versions and only differs in the portion of the total distribution function. 
However, one can now observe differences not only in the species-specific relaxation of the heat flux, which are even more pronounced in this case, but also in the mixture heat flux relaxation, although the relative error remains very small for all three approaches. 
Similar to instance 1, the empirical relaxation frequency $\nu\spec{\alpha}_{\mathrm{empi}}$ shows the best overall result.

\subsubsection{Case 2}
In Case 2, the problem becomes more complicated due to three species being used, whereby the mass of He is significantly smaller than the masses of the other two species.
In addition, there is a density difference between all species and one initial temperature outside the thermal equilibrium, as shown in Table~\ref{tab:initcase2}.
The particles are initialized with a Maxwell distribution function.
\begin{table}[ht]
\centering
\begin{tabular}{|c|c|c|c|}
\hline
\textbf{Species} & \({n_{\text{init}}}\) / m$^{-3}$& \({T_{\text{init}}}\) / K& \({\mathbf u_{\text{init}}}\) / \si{\meter\per\second} \\ 
\hline
Argon Ar & $2\cdot 10^{22}$ & 10000 & $(0,0,0)^\text{T}$ \\ 
\hline
Nitrogen N & $2\cdot 10^{21}$ & 5000 & $(0,0,0)^\text{T}$ \\ 
\hline
Helium He & $8\cdot 10^{21}$ & 5000 & $(0,0,0)^\text{T}$ \\ 
\hline
\end{tabular}
\caption{Initial parameters for 0D reservoir simulation, case 2.\label{tab:initcase2}}
\end{table}

\begin{figure}\centering
 \includegraphics[width=0.6\linewidth]{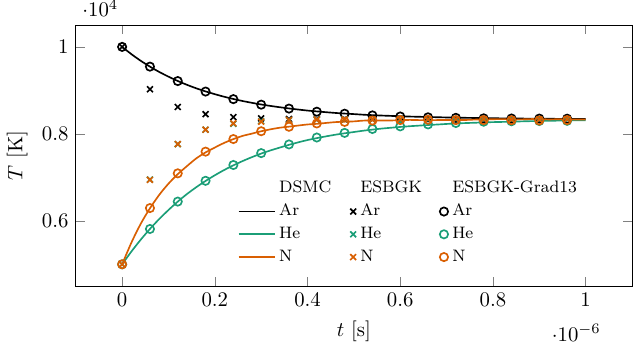}
  \caption{Species temperature relaxation with DSMC, ESBGK and the newly proposed ESBGK mixture model 0D reservoir simulation, case 2.}\label{fig:homrel3spectemp}
\end{figure}

The temperature relaxation is depicted in Fig.~\ref{fig:homrel3spectemp}.
There is a very good agreement between the presented ESBGK model and the temperature relaxation in DSMC.
The results for all three relaxation frequencies of the new ESBGK model are again identical, as expected.
However, there is a strong deviation of the ESBGK mixture model from~\cite{bgk-multispecies}.
In particular, it can be seen that apart from a fundamentally incorrect relaxation time, a separation between N and He cannot be represented in the old ESBGK model, as there is only one relaxation term and thus only one relaxation time for correcting the mixture heat flux relaxation.
In the model presented, however, the correct relaxation of the temperatures is achieved, because the different relaxation times are included directly in the $\hat{T}\spec{\alpha,\text{rel}}$ introduced here.

\subsubsection{Case 3}
In Case 3, Ar and He are considered as a mixture of two species, which again results in a large difference in mass.
The simulation is initialized in thermal equilibrium with a Maxwell distribution function.
However, both species are initialized with opposite velocities, as indicated in Table~\ref{tab:initcase3}.
\begin{table}[ht]
\centering
\begin{tabular}{|c|c|c|c|}
\hline
\textbf{Species} & \({n_{\text{init}}}\) / m$^{-3}$& \({T_{\text{init}}}\) / K & \({\mathbf u_{\text{init}}}\) / \si{\meter\per\second} \\ 
\hline
Argon Ar & $2\cdot 10^{22}$ & 5000 & $(-1000,0,0)^\text{T}$ \\ 
\hline
Helium He & $6\cdot 10^{21}$ & 5000 & $(1000,0,0)^\text{T}$ \\ 
\hline
\end{tabular}
\caption{Initial parameters for 0D reservoir simulation, case 3.\label{tab:initcase3}}
\end{table}

\begin{figure}\centering
  \subfloat[Temperature]{\includegraphics[width=0.6\linewidth]{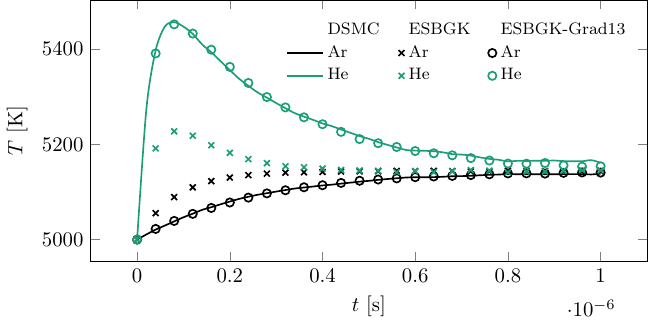}}\\
  \subfloat[Velocity]{\includegraphics[width=0.6\linewidth]{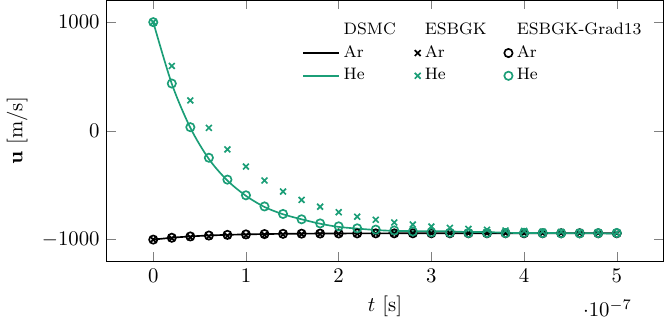}}
  \caption{Temperature and velocity relaxation with DSMC, ESBGK and the newly proposed ESBGK mixture model 0D reservoir simulation, case 3.}\label{fig:homrel2specvelodiff}
\end{figure}

\begin{figure}\centering
  \includegraphics[width=0.6\linewidth]{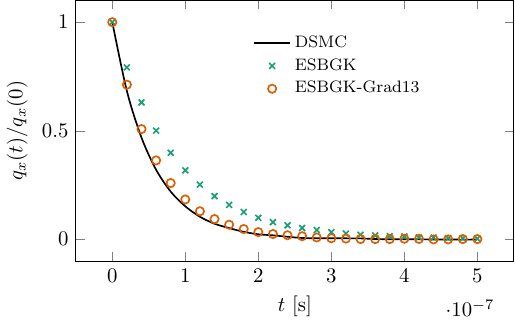}
  \caption{Heat flux relaxation with DSMC, ESBGK and the newly proposed ESBGK mixture model 0D reservoir simulation, case 3.}\label{fig:homrel2specvelodiffheat}
\end{figure}

In this case, the different starting velocities lead to an increase in the species temperatures during relaxation to a common velocity.
With the presented ESBGK model, this process can also be modeled with the correct rate in the velocity and the temperature relaxation for each species, as shown in Fig.~\ref{fig:homrel2specvelodiff}.
Due to the different starting velocities of the species, the overall distribution function of the mixture shows a skewness, i.e. a resulting heat flux, although both species are generated independently with a Maxwell distribution.
The new ESBGK model also matches the heat flux relaxation of the reference DSMC results very well, as shown in Fig.~\ref{fig:homrel2specvelodiffheat}.
The correct relaxation of the temperatures and velocities reduces the error in the relaxation of the heat flux of the mixture.
Also, the heat flux relaxation in this case is again identical for all three relaxation frequencies, and therefore, the choice does not play a significant role.

\subsubsection{Case 4}
In the fourth case, a mixture consisting of four species is used.
A strong non-equilibrium is generated in the initialization.
For this, the Grad 13 equation~\eqref{eq:grad13} is again used to initialize the mixture with a heat flux and pressure tensor.
In addition, the starting temperatures and velocities of each species are different as shown in Table~\ref{tab:initcase4}.
Only He and O have the same starting temperature to investigate the separation of the temperatures.
\begin{table}[ht]
\centering
\begin{tabular}{|c|c|c|c|}
\hline
\textbf{Species} & \({n_{\text{init}}}\) / m$^{-3}$ & \({T_{\text{init}}}\) / K & \({\mathbf u_{\text{init}}}\) / \si{\meter\per\second} \\ 
\hline
Argon Ar & $2\cdot 10^{22}$ & 10000 & $(-1000,0,0)^\text{T}$ \\ 
\hline
Nitrogen N & $2\cdot 10^{21}$ & 5000 & $(500,0,0)^\text{T}$ \\ 
\hline
Helium He & $2\cdot 10^{21}$ & 15000 & $(0,0,0)^\text{T}$ \\ 
\hline
Oxygen O & $6\cdot 10^{21}$ & 15000 & $(1000,0,0)^\text{T}$ \\ 
\hline
\end{tabular}
\caption{Initial parameters for 0D reservoir simulation, case 4.\label{tab:initcase4}}
\end{table}

\begin{figure}\centering
  \subfloat[Pressure tensor]{\includegraphics[width=0.35\linewidth]{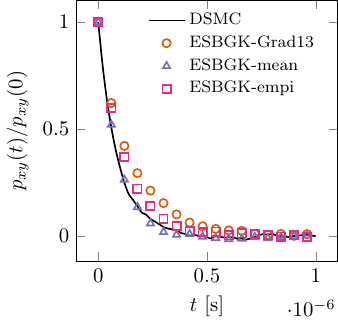}}\quad
  \subfloat[Heat flux]{\includegraphics[width=0.35\linewidth]{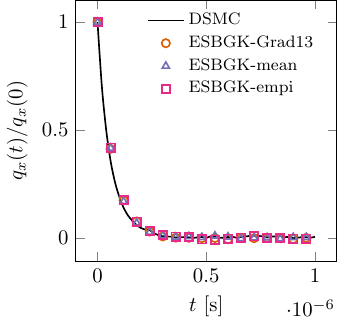}}
  \caption{Pressure tensor and heat flux relaxation with DSMC, ESBGK and the newly proposed ESBGK mixture model 0D reservoir simulation, case 4.}\label{fig:homreledspbgklincase4}
\end{figure}

The relaxation of the pressure tensor and the heat flux of the mixture presented in Fig.~\ref{fig:homreledspbgklincase4} shows that the newly proposed ESBGK model with all relaxation frequency variations match the relaxation curves of the DSMC method relatively well.
Thus the correct Prandtl number is achieved.
Here, however, one can now observe a larger deviation of the pressure tensor for $\nu_\text{Grad13}$.
In this case, the problem described in Section~\ref{sec:limitsmet} occurs:
the relaxation frequency becomes very small for some species, which causes the production terms of the pressure tensor to become very large since these typically relax faster than temperatures and velocities, leading to errors.

\begin{figure}\centering
  \subfloat[Temperature]{\includegraphics[width=0.6\linewidth]{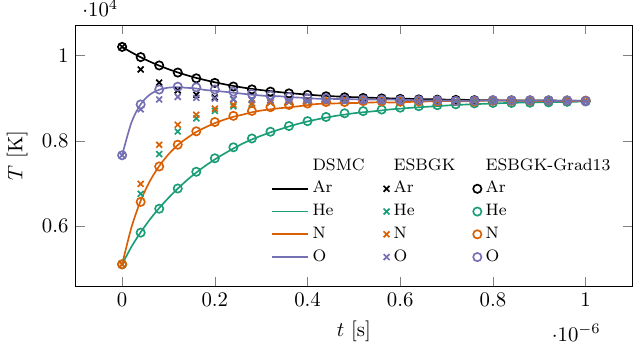}}\\
  \subfloat[Velocity]{\includegraphics[width=0.6\linewidth]{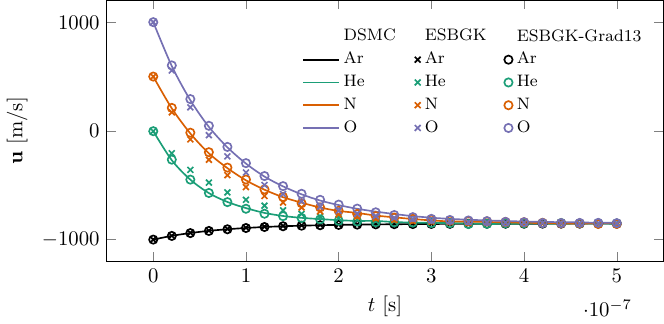}}
  \caption{Temperature and velocity relaxation with DSMC, ESBGK and the newly proposed ESBGK mixture model 0D reservoir simulation, case 4.}\label{fig:homrel4specvelodiff}
\end{figure}

In the relaxation of the temperature and velocities, there are major differences between the old ESBGK model and the one presented here, which perfectly matches the DSMC results (independent of the chosen relaxation frequency) as depicted in Fig.~\ref{fig:homrel4specvelodiff}.
In particular, the clear temperature separation between He and N is virtually not represented by the old ESBGK model and is perfectly met by the model presented.

\subsection{Mass Diffusion}
\label{sec:massdiff}
The next test case is a mass diffusion.
Here, the right and left sides of the simulation domain serve as reservoirs for different species. 
The 1D computational domain is $4\,\mathrm{\mu m}$ long, and the gases have a temperature of $T = 273.15\,\mathrm{K}$ with no flow velocity at the boundaries.
The development of the flow is directly dependent on the correct diffusion coefficient. 
For the first two test cases, Ar and He are used as the gas mixture, while the last simulation is performed with a mixture of Ar, He, and atomic nitrogen N. 
The density conditions in the reservoirs for the different cases are shown in Table~\ref{tab:massdifcon}.
\begin{table}[h!]
\centering
\begin{tabular}{lccc}
\hline
\textbf{Case} & Ar / m$^{-3}$ & He / m$^{-3}$ & N / m$^{-3}$ \\
\hline
1 & $5.37332\cdot 10^{24}$ & $5.37332\cdot 10^{24}$ & -- \\
2 & $5.37332\cdot 10^{25}$ & $5.37332\cdot 10^{25}$ & -- \\
3 & $5.37332\cdot 10^{24}$ & $5.37332\cdot 10^{24}$ & $5.37332\cdot 10^{24}$ \\
\hline
\end{tabular}
\caption{Reservoir densities for the mass diffusion test cases. In Case 3, He and N are located on the same reservoir side.}
\label{tab:massdifcon}
\end{table}

\subsubsection{Mass Diffusion: Case 1}
\begin{figure}\centering
  \includegraphics[width=0.7\linewidth]{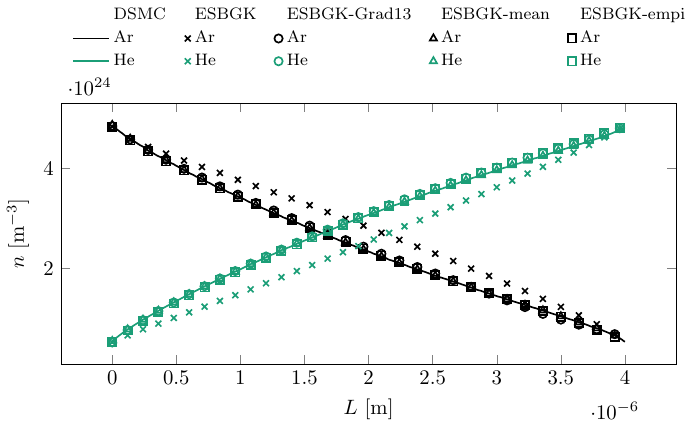}
  \caption{Stationary density profiles of the species for Case~1 of the mass diffusion.}\label{fig:massdiff1}
\end{figure}

The results of the stationary density profiles of the species for Case~1 are shown in Fig.~\ref{fig:massdiff1}. 
For this case, a very good agreement between the proposed model and the DSMC results can be observed, while the expected deviations of the old ESBGK model appear due to its inability to reproduce Fickian diffusion. 
In this case, no differences can be seen between the three relaxation frequencies.

\subsubsection{Mass Diffusion: Case 2}
\begin{figure}\centering
  \includegraphics[width=0.7\linewidth]{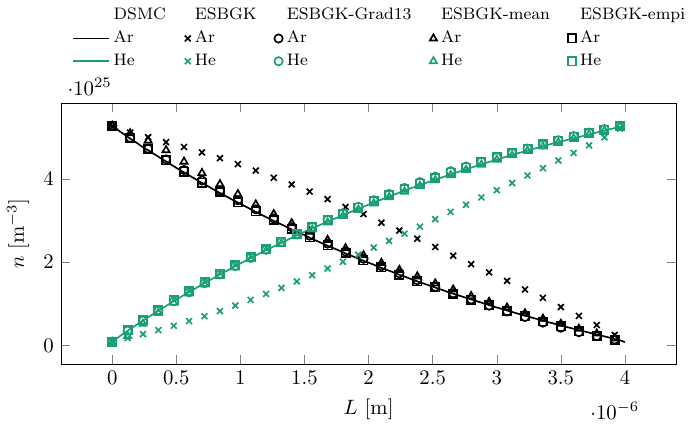}
  \caption{Stationary density profiles of the species for Case~2 of the mass diffusion.}\label{fig:massdiff2}
\end{figure}

Case~2 is identical to Case~1 except from a ten times larger density.
As a result, the error of the old ESBGK model becomes significantly larger. 
The newly proposed model again matches the DSMC results very well. 
It should be noted that the relaxation frequencies $\nu_\text{Grad13}$ and $\nu_\text{empi}$ agree almost perfectly with DSMC, 
while the mean relaxation frequency $\nu_\text{mean}$ shows a slight deviation for Ar at the left boundary as shown in Fig.~\ref{fig:massdiff2}. 
This is because of the situation described in Section~\ref{sec:limitsmet}:
Due to very large relative velocities, $T\spec{\alpha,\mathrm{rel}}$ becomes negative in one cell. 
The deviation from the DSMC result is then caused by the fallback mechanism described earlier, $\mathbf{u}^{(\alpha)}_{\mathrm{rel}} = \mathbf{u}$.

\subsubsection{Mass Diffusion: Case 3}
\begin{figure}\centering
  \includegraphics[width=0.7\linewidth]{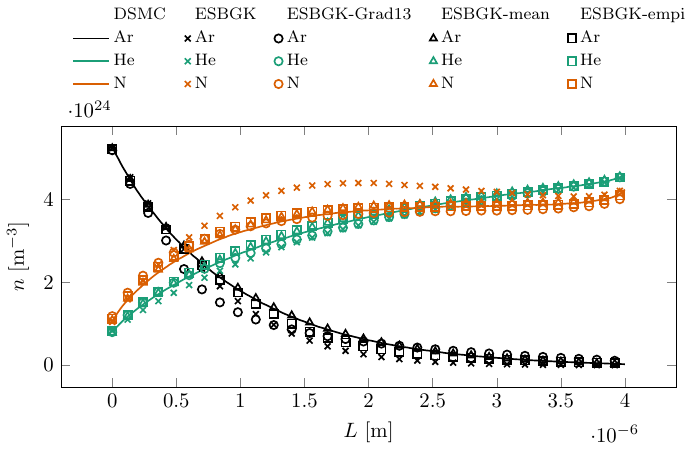}
  \caption{Stationary density profiles of the species for Case~3 of the mass diffusion.}\label{fig:massdiff3}
\end{figure}

The third case is more complex due to three species being involved. 
Here, a large deviation for N is observed when using the old ESBGK model, while the new model again matches the DSMC results very well, which is shown in Fig.~\ref{fig:massdiff3}. 
This time, a larger deviation can be seen for the Grad 13 relaxation frequency $\nu_\text{Grad13}$ at $L \approx 0.75\times 10^{-6}\,\mathrm{m}$ for Ar. 
The reason is the same as in Case~2 for the mean relaxation frequency $\nu_\text{mean}$:
$T\spec{\alpha,\mathrm{rel}}$ becomes negative in two cells, which leads to the deviation due to the fallback mechanism. 
In summary, for all mass diffusion test cases, the empirical approach for the relaxation frequency achieves the best overall performance.

\subsection{Supersonic Couette Flow}
\label{sec:supercouette}
The next test case is a supersonic Couette flow to test the limits of the new method. 
The configuration is one-dimensional, with a distance of 1~m between the moving walls, discretized into 100 cells.
The upper and lower walls move with velocities of $v_{\mathrm{top}} = 500\,\si{\meter\per\second}$ and 
$v_{\mathrm{bot}} = -500\,\si{\meter\per\second}$, respectively. 
The gas mixture is initialized at $v_0 = 0\,\si{\meter\per\second}$, $T_0 =273\,\mathrm{K}$, and each species has a particle density of 
$6.5\cdot 10^{19}\,\mathrm{m^{-3}}$.
\subsubsection{N--O Case}
\begin{figure}\centering
  \includegraphics[width=0.5\linewidth]{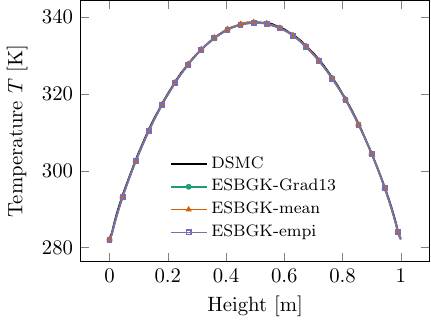}
  \caption{Stationary temperature profile of the N--O mixture Couette flow.}\label{fig:couetteNO}
\end{figure}

In the first case, an N--O mixture is simulated, and the temperature profile depicted in Fig.~\ref{fig:couetteNO} shows a perfect agreement between the results of the DSMC methods and the proposed ESBGK model with all three versions for the relaxation frequency.

\subsubsection{Ar--He Case}
\begin{figure}\centering
  \includegraphics[width=0.5\linewidth]{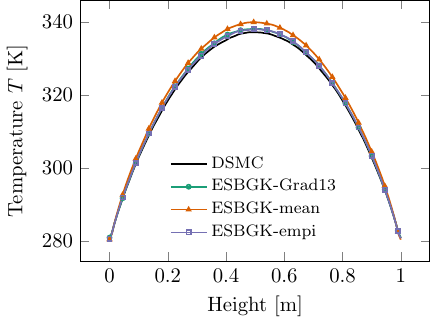}
  \caption{Stationary temperature profile of the Ar-He mixture Couette flow.}\label{fig:couetteAr}
\end{figure}

In the second case, an Ar--He mixture is simulated.
This case is significantly more challenging due to the large mass difference between the two species. 
As shown in Fig.~\ref{fig:couetteAr}, the agreement with DSMC is again very good, even for this more complex scenario. 
However, some differences become visible in this case:
While the Grad 13 and the empirical approaches for the relaxation frequency reproduce the DSMC result almost perfectly, a larger deviation can be observed for the mean relaxation frequency here.

\subsection{70 Degree Blunted Cone}
\begin{figure}[!h]\centering
  \includegraphics[width=0.5\linewidth]{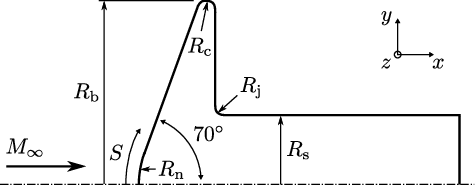}
  \caption{Geometry of the $70^\circ$ blunted cone. $R_{\mathrm{b}}=\SI{25.0}{\milli\meter}$, $R_{\mathrm{c}}=\SI{1.25}{\milli\meter}$, $R_{\mathrm{j}}=\SI{2.08}{\milli\meter}$, $R_{\mathrm{n}}=\SI{12.5}{\milli\meter}$, $R_{\mathrm{s}}=\SI{6.25}{\milli\meter}$. $S$ denotes the arc length along the surface.}\label{fig:70cone_geometry}
\end{figure}

The final verification case considers a hypersonic flow around a $70^\circ$ blunted cone. 
The model geometry, which is based on a wind-tunnel experiment, is illustrated in Fig.~\ref{fig:70cone_geometry}. 
Axisymmetric simulations are carried out with a particle weighting factor that increases in the $y$-direction. 
The surface of the blunted cone is modeled as diffusely reflecting with full thermal accommodation at a constant wall temperature of 
$T_{\mathrm{w}}=300\,\mathrm{K}$. 
Three test cases are investigated using the inflow conditions summarized in Table~\ref{tab:cases}, allowing the influence of different gas compositions and atomic mass ratios to be examined.
\begin{table}[!h]
  \centering
  \caption{Inflow conditions for $70^{\circ}$ blunted cone cases.}
  \label{tab:cases}
  \begin{tabular}{|c|c|c|c|c|}
     \hline
     \textbf{Case} & $n_{\infty}$ / m$^{-3}$ & $T_{\infty}$ / K & $u_{\infty}$ / \si{\meter\per\second} & Composition\\
     \hline
     1 & $7.43\cdot 10^{20}$ & 13.3 & 1502.57 & $\nicefrac{3}{4}$ N -- $\nicefrac{1}{4}$ O \\
     \hline
     2 & $7.43\cdot 10^{20}$ & 13.3 & 1502.57 & $\nicefrac{3}{4}$ Ar -- $\nicefrac{1}{4}$ He \\
     \hline
     3 & $9.29\cdot 10^{20}$ & 13.3 & 1502.57 & $\nicefrac{3}{5}$ Ar -- $\nicefrac{1}{5}$ He -- $\nicefrac{1}{5}$ N \\
     \hline
  \end{tabular}
\end{table}

\subsubsection{Case 1} 
\begin{figure}
\centering
\includegraphics[width=0.6\linewidth]{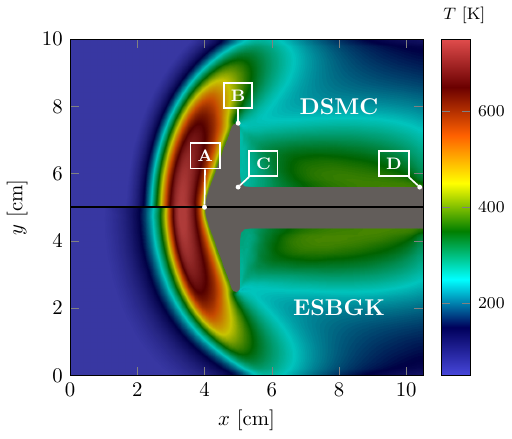}
\caption{$70^\circ$ blunted cone, Case 1: Temperature plots of the flow field using the DSMC method and the proposed ESBGK model.}
\label{Fig:compTtrans}
\end{figure}

\begin{figure}
\centering
\includegraphics[width=0.5\linewidth]{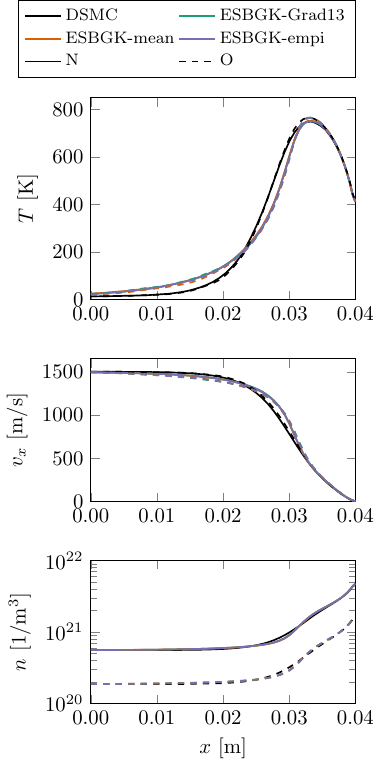}
\caption{$70^\circ$ blunted cone, Case 1: Species temperatures, velocities in $x$ direction, and number densities along the stagnation stream line using DSMC and proposed ESBGK model.}
\label{Fig:set1comp}
\end{figure}

\begin{figure}
\centering
\includegraphics[width=0.5\linewidth]{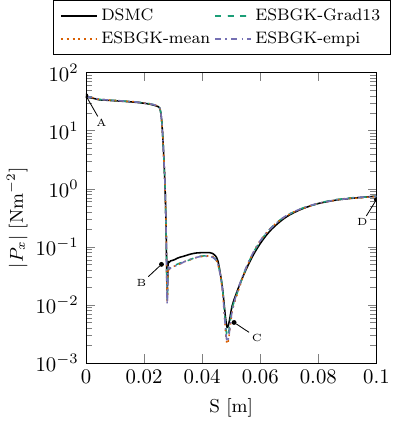}
\caption{$70^\circ$ blunted cone, Case 1: Pressure in $x$ direction on the surface.}
\label{Fig:set1heat}
\end{figure}

\begin{figure}
\centering
\includegraphics[width=0.5\linewidth]{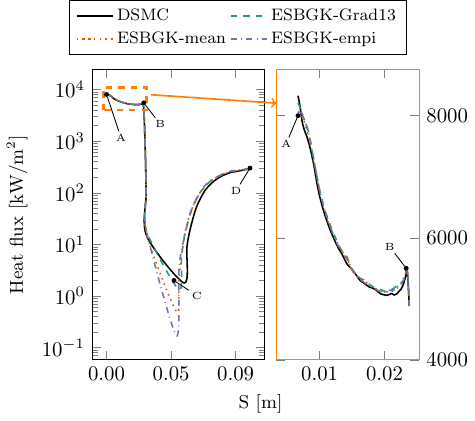}
\caption{$70^\circ$ blunted cone, Case 1: Heat flux on the surface.\label{fig:forcecase1}}
\end{figure}

The first test is performed with an atomic nitrogen-oxygen mixture at a $75\,\%$~-~$25\,\%$ ratio.
A comparison of the mean translational temperature of the gas mixture using DSMC and the new ESBGK model is depicted in Fig.~\ref{Fig:compTtrans}.
The results using the three different relaxation frequency definitions are almost identical in this case.
The ESBGK model predicts an early onset of the temperature increase compared to DSMC, resulting in slightly wider shock profiles.
However, the overall agreement with the DSMC results is very good.
The simulation results of the species flow variables on the stagnation stream line are shown in Fig.~\ref{Fig:set1comp}.
The overall agreement for all relaxation frequency models with the DSMC result is very good.
Again, small differences in the temperature can be observed during the onset of the shock, however, the agreement in the post-shock region is excellent. 
The heat flux and pressure on the cone surface are depicted in Fig.~\ref{Fig:set1heat} and Fig.~\ref{fig:forcecase1} with the points \{A,B,C,D\} corresponding to the points depicted in Fig.~\ref{Fig:compTtrans}.
Both show excellent agreement on the flow-facing heat shield as well backside of the cone further downstream.

\subsubsection{Case 2} 
\begin{figure}
\centering
\includegraphics[width=0.5\linewidth]{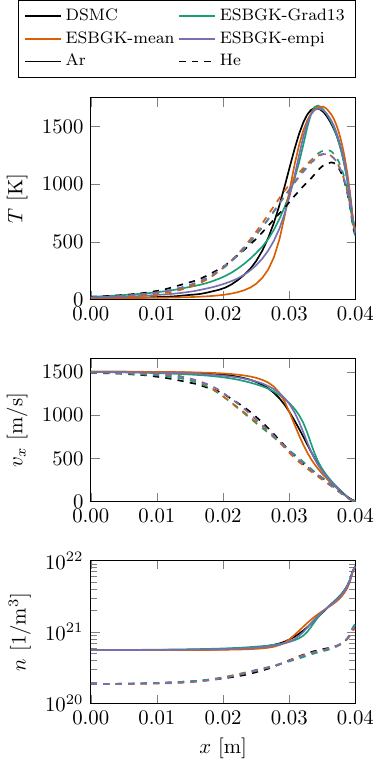}
\caption{$70^\circ$ blunted cone, Case 2: Species temperatures, velocities in $x$ direction, and number densities along the stagnation stream line using DSMC and proposed ESBGK model.}
\label{Fig:set2comp}
\end{figure}

\begin{figure}
\centering
\includegraphics[width=0.5\linewidth]{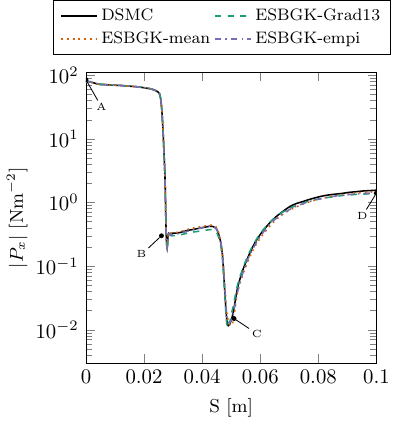}
\caption{$70^\circ$ blunted cone, Case 2: Pressure in $x$ direction on the surface.\label{fig:forcecase2}}
\end{figure}

\begin{figure}
\centering
\includegraphics[width=0.5\linewidth]{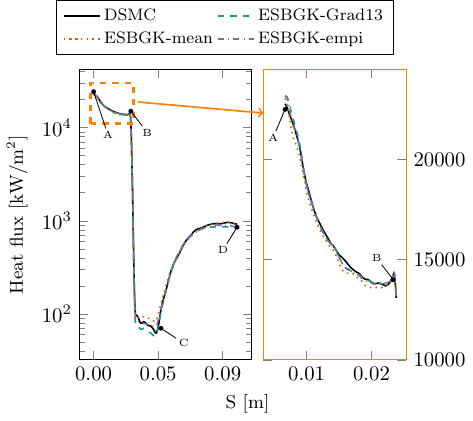}
\caption{$70^\circ$ blunted cone, Case 2: Heat flux on the surface.}
\label{Fig:set2heat}
\end{figure}

Case 2 is performed with an argon-helium mixture at a $75\,\%$~-~$25\,\%$ ratio.
This mixture has a significantly higher mass ratio $\nicefrac{m_{\mathrm{Ar}}}{m_{\mathrm{He}}}\approx 10$ compared to Case 1 with $\nicefrac{m_{\mathrm{O}}}{m_{\mathrm{N}}}\approx 1.14$.
Therefore, a greater difference between the three relaxation frequency versions is expected here. 
The results of the species flow variables on the stagnation stream line are depicted in Fig.~\ref{Fig:set2comp}.
Overall, the agreement between the ESBGK models and the DSMC results is very good, even though a pronounced thermal non-equilibrium develops between Ar and He and the species velocities differ significantly. 
In particular, for Ar, noticeable differences are observed among the three relaxation frequency models. 
Among these, the empirical model exhibits the best agreement with the DSMC simulations.
The pressure on the surface depicted in Fig.~\ref{fig:forcecase2} as well as the heat flux shown in Fig.~\ref{Fig:set2heat} are almost identical for DSMC and the different relaxation frequency models.
The Grad 13 and empirical models exhibit a slightly better agreement than the mean model for the heat flux, but the differences are marginal.

\subsubsection{Case 3} 
\begin{figure}
\centering
\includegraphics[width=0.5\linewidth]{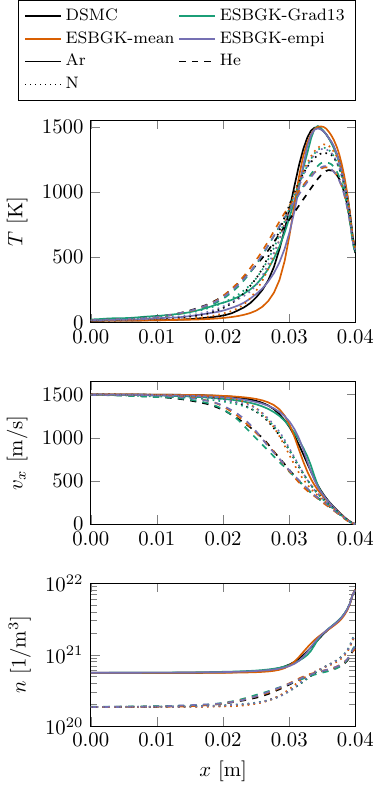}
\caption{$70^\circ$ blunted cone, Case 3: Species temperatures, velocities in $x$ direction, and number densities along the stagnation stream line using DSMC and proposed ESBGK model.}
\label{fig:set3comp}
\end{figure}

\begin{figure}
\centering
\includegraphics[width=0.5\linewidth]{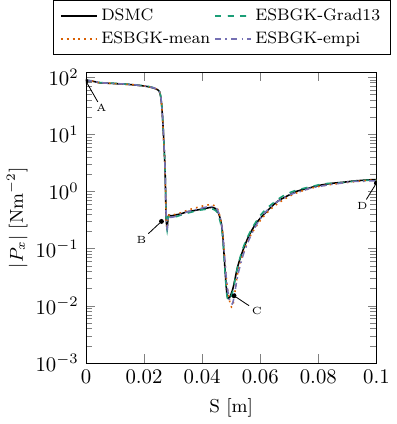}
\caption{$70^\circ$ blunted cone, Case 3: Pressure in $x$ direction on the surface.\label{fig:forcecase3}}
\end{figure}

\begin{figure}
\centering
\includegraphics[width=0.5\linewidth]{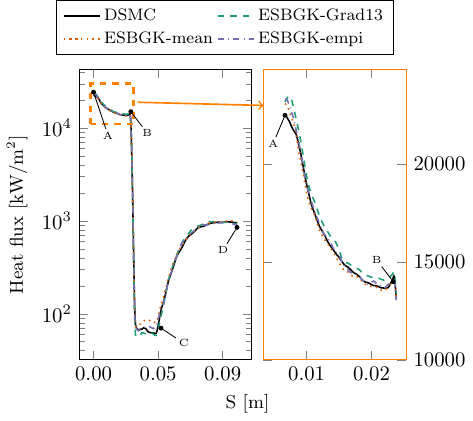}
\caption{$70^\circ$ blunted cone, Case 3: Heat flux on the surface.}
\label{fig:set3heat}
\end{figure}

The third case is performed with an argon-helium-nitrogen mixture.
The three-species mixture and the large mass differences further increase the complexity of the test case.
Nevertheless, the species flow variables along the stagnation streamline (Fig.~\ref{fig:set3comp}) show very good agreement between the DSMC results and all ESBGK models.
Once again, the largest differences among the relaxation frequency models are observed in the temperature profiles of Ar, which has the largest molecular mass.
Here, the empirical model shows again the most pronounced agreement with the DSMC results. 
It should be noted, however, that the strongly varying density and velocity profiles of the different species are captured very well by all the relaxation models.
The pressure on the surface depicted in Fig.~\ref{fig:forcecase3} as well as the heat flux shown in Fig.~\ref{fig:set3heat} are almost identical again for DSMC and ESBGK.
In this case, only a minor deviation is observed for the Grad 13 relaxation model in the heat flux. 
Overall, the empirical model performs best for all simulated $70^\circ$ blunted cone cases.

\section{Conclusion}
\label{sec:conclusion}
A new multi-species ESBGK model for gas mixtures has been presented that correctly reproduces the relaxation of species velocities, temperatures, and pressure tensors according to the Boltzmann collision integral, while maintaining the correct Prandtl number of the mixture.
The central idea is the introduction of species-specific relative relaxation targets, a relative temperature, a relative flow velocity, and a relative pressure tensor per species, which encode the correct inter-species exchange rates derived from the Grad 13 approximation for VHS gases.
Importantly, this is achieved with only a single relaxation term per species, in contrast to models that require $N$ relaxation operators per species for a mixture of $N$ components.
A key advantage of this approach is that despite its relative simplicity, the model captures physical phenomena that are fundamentally inaccessible to prior single-term ESBGK mixture models:
The correct species temperature separation, the correct species velocity relaxation rates, and the correct species pressure tensor relaxation are simultaneously achieved.
In particular, verification cases demonstrate that a former presented ESBGK model~\cite{bgk-multispecies} with a single mixture relaxation term fails to solve the temperature separation between species of very different mass, whereas the proposed model matches the DSMC reference results with high accuracy across all tested configurations, including binary and ternary mixtures with mass ratios up to $m_\mathrm{Ar}/m_\mathrm{He} \approx 10$.
Three variants of the relaxation frequency have been proposed and assessed.
The Grad 13 frequency $\nu_\mathrm{Grad13}$ and the mixture-mean frequency $\nu_\mathrm{mean}$ exhibit complementary over- and underestimation of the species-wise heat flux relaxation, while the empirical harmonic mean $\nu_\mathrm{empi}$ consistently yields the best overall agreement with DSMC across all test cases, including the challenging hypersonic flow around a 70$^\circ$ blunted cone with ternary gas mixtures.
Since the model requires only one relaxation operator per species rather than $N$, and since only the moments of the individual species distributions, which are required anyway for computing the inter-species exchange rates, are needed, the additional computational overhead compared to the previous single-term ESBGK model is modest.
The formulation is straightforward to implement in both particle-based methods and deterministic Discrete Velocity Method (DVM) solvers, as the modifications relative to a standard ESBGK implementation reduce to the computation of the per-species relative relaxation targets and the chosen relaxation frequency.
The model therefore represents a computationally efficient extension of existing ESBGK frameworks that substantially improves accuracy in thermal and velocity non-equilibrium without sacrificing the structural simplicity that makes BGK-type methods attractive for multi-scale gas flow simulations.
\section*{Acknowledgments}
M. Pfeiffer's and F. Tuttas's work is funded by the Deutsche Forschungsgemeinschaft (DFG, German Research Foundation) -- Project-ID 516238647 -- SFB 1667/1 (ATLAS -- Advancing Technologies of Very Low-Altitude Satellites)
\bibliographystyle{plainnat}
\bibliography{sample}

\begin{thebibliography}{51}
\providecommand{\natexlab}[1]{#1}
\providecommand{\url}[1]{\texttt{#1}}
\expandafter\ifx\csname urlstyle\endcsname\relax
  \providecommand{\doi}[1]{doi: #1}\else
  \providecommand{\doi}{doi: \begingroup \urlstyle{rm}\Url}\fi

\bibitem[Andries et~al.(2002)Andries, Aoki, and Perthame]{andries2002consistent}
Pierre Andries, Kazuo Aoki, and Benoit Perthame.
\newblock A consistent bgk-type model for gas mixtures.
\newblock \emph{Journal of Statistical Physics}, 106:\penalty0 993--1018, 2002.

\bibitem[Asinari(2008)]{asinari2008asymptotic}
Pietro Asinari.
\newblock Asymptotic analysis of multiple-relaxation-time lattice boltzmann schemes for mixture modeling.
\newblock \emph{Computers \& Mathematics with Applications}, 55\penalty0 (7):\penalty0 1392--1407, 2008.

\bibitem[Bhatnagar et~al.(1954)Bhatnagar, Gross, and Krook]{bgk}
P.~L. Bhatnagar, E.~P. Gross, and M.~Krook.
\newblock {A Model for Collision Processes in Gases. I. Small Amplitude Processes in Charged and Neutral One-Component Systems}.
\newblock \emph{Phys. Rev.}, 94:\penalty0 511--525, 1954.
\newblock \doi{10.1103/PhysRev.94.511}.

\bibitem[Bird(1994)]{bird}
G.~A. Bird.
\newblock \emph{{Molecular Gas Dynamics and the Direct Simulation of Gas Flows}}.
\newblock Oxford University Press, New York, 1994.

\bibitem[Bisi et~al.(2024)Bisi, Groppi, Lucchin, Martalo, et~al.]{bisi2024mixed}
Marzia Bisi, Maria Groppi, Enrico Lucchin, Giorgio Martalo, et~al.
\newblock A mixed boltzmann--bgk model for inert gas mixtures.
\newblock \emph{Kinetic and Related Models}, 17\penalty0 (5):\penalty0 674--696, 2024.

\bibitem[Bobylev et~al.(2018)Bobylev, Bisi, Groppi, Spiga, and Potapenko]{bobylev2018general}
Alexander~V Bobylev, Marzia Bisi, Maria Groppi, Giampiero Spiga, and Irina~F Potapenko.
\newblock A general consistent bgk model for gas mixtures.
\newblock \emph{Kinetic \& Related Models}, 11\penalty0 (6), 2018.

\bibitem[Brull(2014)]{brull}
S.~Brull.
\newblock {An ellipsoidal statistical model for gas mixtures}.
\newblock \emph{Commun. Math. Sci.}, 13:\penalty0 1--13, 2014.
\newblock \doi{10.4310/CMS.2015.v13.n1.a1}.

\bibitem[Brull(2021)]{brull2021ellipsoidal}
S.~Brull.
\newblock An ellipsoidal statistical model for a monoatomic and a polyatomic gas mixture.
\newblock \emph{Commun. Math. Sci.}, 19\penalty0 (8):\penalty0 2177--2194, 2021.

\bibitem[Burt and Boyd(2008)]{burt-boyd-ld}
J.~M. Burt and I.~D. Boyd.
\newblock {A low diffusion particle method for simulating compressible inviscid flows}.
\newblock \emph{J. Comput. Phys.}, 227\penalty0 (9):\penalty0 4653--4670, 2008.
\newblock \doi{10.1016/j.jcp.2008.01.020}.

\bibitem[Burt and Boyd(2006)]{burt2006evaluation}
Jonathan Burt and Iain Boyd.
\newblock Evaluation of a particle method for the ellipsoidal statistical bhatnagar-gross-krook equation.
\newblock In \emph{44th AIAA aerospace sciences meeting and exhibit}, page 989, 2006.

\bibitem[Fasoulas et~al.(2019)Fasoulas, Munz, Pfeiffer, Beyer, Binder, Copplestone, Mirza, Nizenkov, Ortwein, and Reschke]{piclas}
S.~Fasoulas, C.-D. Munz, M.~Pfeiffer, J.~Beyer, T.~Binder, S.~Copplestone, A.~Mirza, P.~Nizenkov, P.~Ortwein, and W.~Reschke.
\newblock {Combining particle-in-cell and direct simulation Monte Carlo for the simulation of reactive plasma flows}.
\newblock \emph{Phys. Fluids}, 31:\penalty0 072006, 2019.
\newblock \doi{10.1063/1.5097638}.

\bibitem[Fei et~al.(2020{\natexlab{a}})Fei, Liu, Liu, and Zhang]{fei2020benchmark}
Fei Fei, Haihong Liu, Zhaohui Liu, and Jun Zhang.
\newblock A benchmark study of kinetic models for shock waves.
\newblock \emph{AIAA Journal}, 58\penalty0 (6):\penalty0 2596--2608, 2020{\natexlab{a}}.

\bibitem[Fei et~al.(2020{\natexlab{b}})Fei, Zhang, Li, and Liu]{FEI2020108972}
Fei Fei, Jun Zhang, Jing Li, and ZhaoHui Liu.
\newblock {A unified stochastic particle Bhatnagar-Gross-Krook method for multiscale gas flows}.
\newblock \emph{J. Comput. Phys.}, 400:\penalty0 108972, 2020{\natexlab{b}}.
\newblock \doi{10.1016/j.jcp.2019.108972}.

\bibitem[Fei et~al.(2021)Fei, Ma, Wu, and Zhang]{fei2021efficient}
Fei Fei, Yang Ma, Jie Wu, and Jun Zhang.
\newblock An efficient algorithm of the unified stochastic particle {{Bhatnagar-Gross-Krook}} method for the simulation of multi-scale gas flows.
\newblock \emph{Advances in Aerodynamics}, 3\penalty0 (1):\penalty0 18, July 2021.
\newblock ISSN 2524-6992.
\newblock \doi{10.1186/s42774-021-00069-8}.

\bibitem[Frolova(2023)]{frolova2023numerical}
Anna~Averkievna Frolova.
\newblock Numerical and theoretical analysis of model equations for multicomponent rarefied gas.
\newblock \emph{Computational Mathematics and Mathematical Physics}, 63\penalty0 (12):\penalty0 2257--2266, 2023.

\bibitem[Gallis and Torczynski(2011)]{gallis-torczynski-2}
M.~A. Gallis and J.~R. Torczynski.
\newblock {Investigation of the ellipsoidal-statistical Bhatnagar–Gross–Krook kinetic model applied to gas-phase transport of heat and tangential momentum between parallel walls}.
\newblock \emph{Phys. Fluids}, 23:\penalty0 030601, 2011.
\newblock \doi{10.1063/1.3558869}.

\bibitem[Garz{\'o} et~al.(1989)Garz{\'o}, Santos, and Brey]{garzo1989kinetic}
Vicente Garz{\'o}, Andres Santos, and J~Javier Brey.
\newblock A kinetic model for a multicomponent gas.
\newblock \emph{Physics of Fluids A: Fluid Dynamics}, 1\penalty0 (2):\penalty0 380--383, 1989.

\bibitem[Gorji and Jenny(2014)]{hossein}
M.~H. Gorji and P.~Jenny.
\newblock {An efficient particle Fokker–Planck algorithm for rarefied gas flows}.
\newblock \emph{J. Comput. Phys.}, 262:\penalty0 325--343, 2014.
\newblock \doi{10.1016/j.jcp.2013.12.046}.

\bibitem[Guo et~al.(2013)Guo, Xu, and Wang]{guo2013discrete}
Zhaoli Guo, Kun Xu, and Ruijie Wang.
\newblock Discrete unified gas kinetic scheme for all {{Knudsen}} number flows: {{Low-speed}} isothermal case.
\newblock \emph{Physical Review E}, 88\penalty0 (3):\penalty0 033305, September 2013.
\newblock ISSN 1539-3755, 1550-2376.
\newblock \doi{10.1103/PhysRevE.88.033305}.

\bibitem[Gupta(2015)]{gupta2015mathematical}
Vinay~Kumar Gupta.
\newblock \emph{Mathematical modeling of rarefied gas mixtures}.
\newblock PhD thesis, Dissertation, Aachen, Techn. Hochsch., 2015, 2015.

\bibitem[Hamel(1965)]{hamel1965kinetic}
Bernard~B Hamel.
\newblock Kinetic model for binary gas mixtures.
\newblock \emph{The Physics of Fluids}, 8\penalty0 (3):\penalty0 418--425, 1965.

\bibitem[Hepp et~al.(2020{\natexlab{a}})Hepp, Grabe, and Hannemann]{hepp2020kinetic}
Christian Hepp, Martin Grabe, and Klaus Hannemann.
\newblock A kinetic fokker--planck approach to model hard-sphere gas mixtures.
\newblock \emph{Physics of Fluids}, 32\penalty0 (2), 2020{\natexlab{a}}.

\bibitem[Hepp et~al.(2020{\natexlab{b}})Hepp, Grabe, and Hannemann]{hepp2020kineticvhs}
Christian Hepp, Martin Grabe, and Klaus Hannemann.
\newblock A kinetic fokker--planck approach for modeling variable hard-sphere gas mixtures.
\newblock \emph{AIP Advances}, 10\penalty0 (8), 2020{\natexlab{b}}.

\bibitem[Hild and Pfeiffer(2024)]{bgk-poly-mix-hild}
F.~Hild and M.~Pfeiffer.
\newblock {Multi-species modeling in the particle-based ellipsoidal statistical Bhatnagar-Gross-Krook method including internal degrees of freedom}.
\newblock \emph{Journal of Computational Physics}, 514:\penalty0 113226, 2024.
\newblock ISSN 0021-9991.
\newblock \doi{https://doi.org/10.1016/j.jcp.2024.113226}.

\bibitem[{Holway Jr.}(1966)]{esbgk}
H.~L. {Holway Jr.}
\newblock {New Statistical Models for Kinetic Theory: Methods of Construction}.
\newblock \emph{Phys. Fluids}, 9:\penalty0 1658--1673, 1966.
\newblock \doi{10.1063/1.1761920}.

\bibitem[Jun et~al.(2019)Jun, Pfeiffer, Mieussens, and Gorji]{Jun2019}
Eunji Jun, Marcel Pfeiffer, Luc Mieussens, and M.~Hossein Gorji.
\newblock {Comparative Study Between Cubic and Ellipsoidal Fokker–Planck Kinetic Models}.
\newblock \emph{AIAA J.}, 57\penalty0 (6):\penalty0 2524--2533, 2019.
\newblock \doi{10.2514/1.J057935}.

\bibitem[Kim et~al.(2026)Kim, Kim, Park, and Jun]{kim2026particle}
Inchan Kim, Joonbeom Kim, Woonghwi Park, and Eunji Jun.
\newblock A particle multi-relaxation bhatnagar-gross-krook method for rarefied monatomic gas mixtures.
\newblock \emph{arXiv preprint arXiv:2604.24244}, 2026.

\bibitem[Kim and Jun(2025{\natexlab{a}})]{kim2025particle}
Sanghun Kim and Eunji Jun.
\newblock A particle fokker--planck method for rarefied gas flows of monatomic mixtures.
\newblock \emph{Physics of Fluids}, 37\penalty0 (1), 2025{\natexlab{a}}.

\bibitem[Kim and Jun(2025{\natexlab{b}})]{kim2025stochastic}
Sanghun Kim and Eunji Jun.
\newblock A stochastic particle method based on the fokker--planck master equation for rarefied gas flows of diatomic mixtures.
\newblock \emph{Physics of Fluids}, 37\penalty0 (3), 2025{\natexlab{b}}.

\bibitem[Klingenberg et~al.(2018{\natexlab{a}})Klingenberg, Pirner, and Puppo]{klingenberg2018consistent}
Christian Klingenberg, Marlies Pirner, and Gabriella Puppo.
\newblock A consistent kinetic model for a two-component mixture with an application to plasma.
\newblock \emph{arXiv preprint arXiv:1806.09462}, 2018{\natexlab{a}}.

\bibitem[Klingenberg et~al.(2018{\natexlab{b}})Klingenberg, Pirner, and Puppo]{klingenberg2018kinetic}
Christian Klingenberg, Marlies Pirner, and Gabriella Puppo.
\newblock Kinetic es-bgk models for a multi-component gas mixture.
\newblock In \emph{Theory, Numerics and Applications of Hyperbolic Problems II: Aachen, Germany, August 2016}, pages 195--208. Springer, 2018{\natexlab{b}}.

\bibitem[Li et~al.(2024)Li, Zeng, and Wu]{li2024kinetic}
Qi~Li, Jianan Zeng, and Lei Wu.
\newblock Kinetic modelling of rarefied gas mixtures with disparate mass in strong non-equilibrium flows.
\newblock \emph{Journal of Fluid Mechanics}, 1001:\penalty0 A5, 2024.

\bibitem[Liu et~al.(2020)Liu, Zhu, and Xu]{liu2020unified}
Chang Liu, Yajun Zhu, and Kun Xu.
\newblock Unified gas-kinetic wave-particle methods {{I}}: {{Continuum}} and rarefied gas flow.
\newblock \emph{Journal of Computational Physics}, 401:\penalty0 108977, January 2020.
\newblock ISSN 0021-9991.
\newblock \doi{10.1016/j.jcp.2019.108977}.

\bibitem[Mathiaud et~al.(2022)Mathiaud, Mieussens, and Pfeiffer]{MATHIAUD202265}
J.~Mathiaud, L.~Mieussens, and M.~Pfeiffer.
\newblock {An ES-BGK model for diatomic gases with correct relaxation rates for internal energies}.
\newblock \emph{Eur. J. Mech. B/Fluids}, 96:\penalty0 65--77, 2022.
\newblock \doi{10.1016/j.euromechflu.2022.07.003}.

\bibitem[Mathiaud and Mieussens(2016)]{mathiaud2016fokker}
Julien Mathiaud and Luc Mieussens.
\newblock A fokker--planck model of the boltzmann equation with correct prandtl number.
\newblock \emph{Journal of Statistical Physics}, 162:\penalty0 397--414, 2016.

\bibitem[Mieussens(2000)]{mieussens2000discrete}
Luc Mieussens.
\newblock Discrete velocity model and implicit scheme for the bgk equation of rarefied gas dynamics.
\newblock \emph{Mathematical Models and Methods in Applied Sciences}, 10\penalty0 (08):\penalty0 1121--1149, November 2000.
\newblock ISSN 0218-2025.
\newblock \doi{10.1142/S0218202500000562}.

\bibitem[Pfeiffer(2018{\natexlab{a}})]{energy-cons}
M.~Pfeiffer.
\newblock {Extending the particle ellipsoidal statistical Bhatnagar-Gross-Krook method to diatomic molecules including quantized vibrational energies}.
\newblock \emph{Phys. Fluids}, 30:\penalty0 116103, 2018{\natexlab{a}}.
\newblock \doi{10.1063/1.5054961}.

\bibitem[Pfeiffer(2018{\natexlab{b}})]{piclas-bgk}
M.~Pfeiffer.
\newblock {Particle-based fluid dynamics: Comparison of different Bhatnagar-Gross-Krook models and the direct simulation Monte Carlo method for hypersonic flows}.
\newblock \emph{Phys. Fluids}, 30:\penalty0 106106, 2018{\natexlab{b}}.
\newblock \doi{10.1063/1.5042016}.

\bibitem[Pfeiffer and Gorji(2017)]{fp}
M.~Pfeiffer and M.~H. Gorji.
\newblock {Adaptive particle–cell algorithm for Fokker–Planck based rarefied gas flow simulations}.
\newblock \emph{Comput. Phys. Commun.}, 213:\penalty0 1--8, 2017.
\newblock \doi{10.1016/j.cpc.2016.11.003}.

\bibitem[Pfeiffer et~al.(2019{\natexlab{a}})Pfeiffer, Mirza, and Nizenkov]{bgk-poly}
M.~Pfeiffer, A.~Mirza, and P.~Nizenkov.
\newblock {Extension of Particle-based BGK Models to Polyatomic Species in Hypersonic Flow around a Flat-faced Cylinder}.
\newblock \emph{AIP Conference Proceedings}, 2132:\penalty0 100001, 2019{\natexlab{a}}.
\newblock \doi{10.1063/1.5119596}.

\bibitem[Pfeiffer et~al.(2019{\natexlab{b}})Pfeiffer, Mirza, and Nizenkov]{pfeiffer2019evaluation}
M~Pfeiffer, A~Mirza, and P~Nizenkov.
\newblock {Evaluation of particle-based continuum methods for a coupling with the direct simulation Monte Carlo method based on a nozzle expansion}.
\newblock \emph{Phys. Fluids}, 31:\penalty0 073601, 2019{\natexlab{b}}.
\newblock \doi{10.1063/1.5098085}.

\bibitem[Pfeiffer et~al.(2021)Pfeiffer, Mirza, and Nizenkov]{bgk-multispecies}
M.~Pfeiffer, A.~Mirza, and P.~Nizenkov.
\newblock {Multi-species modeling in the particle-based ellipsoidal statistical Bhatnagar–Gross–Krook method for monatomic gas species}.
\newblock \emph{Phys. Fluids}, 33:\penalty0 036106, 2021.
\newblock \doi{10.1063/5.0037915}.

\bibitem[Pfeiffer et~al.(2022)Pfeiffer, Garmirian, and Gorji]{PhysRevE.106.025303}
M.~Pfeiffer, F.~Garmirian, and M.~H. Gorji.
\newblock {Exponential Bhatnagar-Gross-Krook integrator for multiscale particle-based kinetic simulations}.
\newblock \emph{Phys. Rev. E}, 106:\penalty0 025303, 2022.
\newblock \doi{10.1103/PhysRevE.106.025303}.

\bibitem[Pfeiffer(2022)]{pfeiffer2022optimized}
Marcel Pfeiffer.
\newblock An optimized collision-averaged variable soft sphere parameter set for air, carbon, and corresponding ionized species.
\newblock \emph{Physics of Fluids}, 34\penalty0 (11), 2022.

\bibitem[Pfeiffer and Tuttas(2026)]{pfeiffer2026shakhovbasedbhatnagargrosskrookmodelpolyatomic}
Marcel Pfeiffer and Franziska Tuttas.
\newblock A shakhov-based bhatnagar-gross-krook model for polyatomic molecules and for atomic as well as polyatomic mixtures, 2026.
\newblock URL \url{https://arxiv.org/abs/2604.01377}.

\bibitem[Pirner(2021)]{overview-bgk-mix}
M.~Pirner.
\newblock {A Review on BGK Models for Gas Mixtures of Mono and Polyatomic Molecules}.
\newblock \emph{Fluids}, 6:\penalty0 393, 2021.
\newblock \doi{10.3390/fluids6110393}.

\bibitem[Schwartzentruber and Boyd(2006)]{SCHWARTZENTRUBER2006402}
T.E. Schwartzentruber and I.D. Boyd.
\newblock A hybrid particle-continuum method applied to shock waves.
\newblock \emph{J. Comput. Phys.}, 215\penalty0 (2):\penalty0 402--416, 2006.
\newblock ISSN 0021-9991.
\newblock \doi{https://doi.org/10.1016/j.jcp.2005.10.023}.

\bibitem[Shakhov(1968)]{shakhov}
E.~M. Shakhov.
\newblock {Generalization of the Krook Kinetic Relaxation Equation}.
\newblock \emph{Fluid Dyn.}, 33:\penalty0 95--96, 1968.
\newblock \doi{10.1007/BF01029546}.

\bibitem[Todorova and Steijl(2019)]{todorova-shakhov-esbgk}
B.~N. Todorova and R.~Steijl.
\newblock {Derivation and numerical comparison of Shakhov and Ellipsoidal Statistical kinetic models for a monoatomic gas mixture}.
\newblock \emph{Eur. J. Mech. B/Fluids}, 76:\penalty0 390--402, 2019.
\newblock \doi{10.1016/j.euromechflu.2019.04.001}.

\bibitem[Todorova et~al.(2020)Todorova, White, and Steijl]{todorova2020}
B.~N. Todorova, C.~White, and R.~Steijl.
\newblock {Modeling of nitrogen and oxygen gas mixture with a novel diatomic kinetic model}.
\newblock \emph{AIP Adv.}, 10:\penalty0 095218, 2020.
\newblock \doi{10.1063/5.0021672}.

\bibitem[Zhang et~al.(2019)Zhang, John, Pfeiffer, Fei, and Wen]{zhang2019particle}
Jun Zhang, Benzi John, Marcel Pfeiffer, Fei Fei, and Dongsheng Wen.
\newblock {Particle-based hybrid and multiscale methods for nonequilibrium gas flows}.
\newblock \emph{Adv. Aerodyn.}, 1:\penalty0 1--15, 2019.
\newblock \doi{10.1186/s42774-019-0014-7}.

\end{thebibliography}

\end{document}